\documentclass[10pt,journal,twoside]{IEEEtran}

\usepackage[english]{babel}
\usepackage{graphicx}
\usepackage{subfigure}
\usepackage{multicol}
\usepackage{setspace}
\usepackage{amsmath}
\usepackage{amsthm}
\usepackage{epstopdf}
\usepackage{fancyhdr}
\usepackage{indentfirst}
\usepackage{enumerate}
\usepackage{caption}
\usepackage{leftidx}
\usepackage{amssymb}
\usepackage{float}
\usepackage{breqn}
\usepackage{color, xcolor}
\usepackage{bm}
\usepackage{cite}
\usepackage{multirow}
\usepackage{totcount}
\usepackage{algorithm}
\usepackage{algcompatible}
\usepackage{hyperref}
\usepackage{bbm}
\usepackage{chngcntr}
\usepackage{setspace}
\counterwithin*{footnote}{page}

\hypersetup{nolinks=true}
\usepackage{booktabs}

\newtheoremstyle{colon}%
{}
{}
{\itshape}
{}
{\itshape}
{:}
{ }
{\thmname{#1}\ \!\thmnumber{\itshape#2}\thmnote{(#3)}}

\theoremstyle{colon}
\newtheorem*{Remark*}{Remark}
\newtheorem*{Theorem*}{Theorem}

\newtheorem*{Lemma*}{Lemma}

\newtheorem{Prop}{Proposition}
\numberwithin{Prop}{section}

\newtheorem{Def}{Definition}
\numberwithin{Def}{section}

\newtheorem{Rem}{Remark}
\numberwithin{Rem}{section}

\newtheorem{Thm}{Theorem}
\numberwithin{Thm}{section}

\numberwithin{Lem}{section}

\DeclareMathOperator*{\argmin}{arg\,min}
\DeclareMathOperator*{\argmax}{arg\,max}

\newcommand{\ts}{\textsuperscript}

\begin{document}
\title{AoI-Delay Tradeoff in Mobile Edge Caching: A Mixed-Order Drift-Plus-Penalty Method}
\author{Ran Li, Chuan Huang, Xiaoqi Qin, and Lei Yang
\thanks{
This work was submitted in part to 2023 IEEE Global Communications Conference Workshops.

R. Li and C. Huang are with the School of Science and Engineering and the Future Network of Intelligence Institute, The Chinese University of Hong Kong, Shenzhen 518172, China (e-mails: ranli2@link.cuhk.edu.cn; huangchuan@cuhk.edu.cn).

X. Qin is with the State Key Laboratory of Networking and Switching Technology, Beijing University of Posts and Telecommunications, Beijing 100876, China (e-mail: xiaoqiqin@bupt.edu.cn).

L. Yang is with the Department of Computer Science and Engineering, University of Nevada, Reno, NV 89557 USA (e-mail: leiy@unr.edu).


}}
\maketitle
\thispagestyle{empty}
\begin{abstract}
Mobile edge caching (MEC) is a promising technique to improve the quality of service (QoS) for mobile users (MU) by bringing data to the network edge. However, optimizing the crucial QoS aspects of message freshness and service promptness, measured by age of information (AoI) and service delay, respectively, entails a tradeoff due to their competition for shared edge resources. This paper investigates this tradeoff by formulating their weighted sum minimization as a sequential decision-making problem, incorporating high-dimensional, discrete-valued, and linearly constrained design variables. First, to assess the feasibility of the considered problem, we characterize the corresponding achievable region by deriving its superset with the rate stability theorem and its subset with a novel stochastic policy, and develop a sufficient condition for the existence of solutions. Next, to efficiently solve this problem, we propose a mixed-order drift-plus-penalty algorithm by jointly considering the linear and quadratic Lyapunov drifts and then optimizing them with dynamic programming (DP). Finally, by leveraging the Lyapunov optimization technique, we demonstrate that the proposed algorithm achieves an $O(1/V)$ versus $O(V)$ tradeoff for the average AoI and average service delay.

\end{abstract}
\section{Introduction}

In recent years, mobile edge caching (MEC) has emerged as a promising solution to tackle the challenges posed by the exponential growth of mobile users (MUs) and the corresponding data demands \cite{ericsson}. To elaborate, MEC enables the base stations (BS) at the network's edge to provide both the uplink and downlink accesses for nearby MUs, which increases the access capacity and network throughput \cite{mec1, mec2,LYang1}. Additionally, MEC deploys caches at the BS to store frequently requested messages, thereby reducing message delivery latency and enhancing the quality of service (QoS) for MUs \cite{app2_4, app2_5}. Due to the aforementioned advantages, MEC has gained significant popularity in various applications, such as Internet of Vehicles (IoV) \cite{mec_veh}, industrial automation networks \cite{mec_ind}, and Internet of Things (IoT) \cite{mec_iot}. For example, in IoV networks \cite{mec_veh}, MEC technique utilizes roadside units or vehicles to access all the surrounding vehicles and cache real-time traffic information. This cached information is fetched by other nearby vehicles to ensure timely route planning and enhance driving safety. Moreover, since the BSs in MEC only need to serve their nearby MUs and can have modest hardware configurations \cite{mec1}, the deployment of MEC generally incurs low infrastructural costs, which makes MEC highly adaptable in more and more applications and positions it to play a pivotal role in shaping the future of mobile networks.

To ensure the efficient adoption of MEC in practical applications, the key issue is to properly schedule limited cache storages and communication resources at the BS to simultaneously meet various QoS demands from the nearby MUs. Specifically, two critical and fundamental QoS demands that need to be addressed across various applications are message freshness and service promptness \cite{app2_4,app2_5,mec_veh,mec_ind,mec_iot}. For instance, message freshness and service promptness are crucial in ensuring safe driving in IoV \cite{mec_veh}, where every vehicle needs to receive timely responses from roadside units regarding the up-to-date information about its surroundings. Additionally, in industrial automation networks \cite{mec_ind}, message freshness and service promptness play a vital role in detecting abnormal situations, where the sensors need to constantly monitor the status of industrial processes and promptly transmit this information to the control center to ensure timely detection and response to emergencies. Message freshness is usually quantified by age of information (AoI) \cite{aoi_original}, which measures the elapsed time since the generation of the MU's previously received message, and service promptness is measured by service delay \cite{mec1}, which captures the time duration from the generation of a request from one MU to the request being served by the BS. Given the significant importance of both the message freshness and service promptness, there is a desire to concurrently attain optimal AoI and service delay in MEC. Unfortunately, achieving this goal is often infeasible, since optimizing either one of these two metrics requires competitively utilization of shared communication resources at the BS \cite{app2_4,app2_5}. Hence, it becomes crucial to thoroughly investigate the relationship between AoI and service delay in MEC and develop scheduling policies that strike an proper balance between them.

\subsection{Related Works}
The existing works on the scheduling problem in MEC cover a wide range of applications, including IoT \cite{app1_task,app1_aod}, IoV \cite{YJZhang1, app1_uav}, non-orthogonal multiple access (NOMA) \cite{app1_fl,app1_noma}, and energy harvesting networks \cite{app1_eh}. These studies primarily addressed the problem of determining ``when to cache" the messages requested by MUs, aiming to optimize AoI or other relevant metrics. In \cite{app1_task}, the authors studied an IoT MEC network with multiple users and multiple edge servers, where the users randomly upload various tasks to the edge servers and the servers utilize shared computation resources to process the uploaded tasks. To optimize the resource utilization, the authors proposed a heuristic resource scheduling policy. In \cite{app1_aod}, the authors defined the age of data (AoD) to measure the quality of big data analytics in IoT MEC networks and proposed a Multi-armed Bandit (MAB) based online learning algorithm to minimize AoD. In \cite{YJZhang1} and \cite{app1_uav}, the authors studied an unmanned aerial vehicle (UAV)-assisted MEC scenario, and addressed the trajectory optimization and computation offloading by using perturbed Lyapunov optimization and successive convex approximation, respectively. In \cite{app1_fl}, the authors leveraged federated learning (FL) in NOMA-based MEC and used graph theory to improve the communication efficiency of FL and to accelerate the model convergence. In \cite{app1_noma}, the authors discussed the power and time allocation in NOMA-assisted MEC and derived the closed-form expression for the optimal MEC offloading policy. In \cite{app1_eh}, the authors focused on the edge resource utility maximization in an energy-harvesting powered MEC network, and proposed a Lyapunov-based algorithm to schedule the edge resources and satisfy the AoI constraints. Although the aforementioned works demonstrated notable AoI improvements for various applications, they primarily scheduled the resources for AoI minimization and neglected to consider the impact of service delay on individual MUs. This oversight may lead to a QoS degradation for MUs, especially in MEC networks with a heavy request load and limited edge resources, where some MUs may never be served. Therefore, the problem of determining ``when to serve" these MUs to effectively reduce the service delay becomes a crucial problem that requires further investigations.

Recently, there has been a surge of interests in investigating the tradeoff between AoI and service delay in MEC networks, considering both the issues of ``when to cache" and ``when to serve". In \cite{app2_1}, the authors studied a MEC system, where one source node (SN) generates time-sensitive messages and only one channel is available for transmitting these messages to the MUs, and characterized the optimal AoI-delay region theoretically. In \cite{app2_2}, the authors also considered the single SN and single channel scenario and derived the closed-form expressions of average AoI and peak AoI (PAoI) to characterize the AoI-delay and PAoI-delay regions, respectively. In \cite{app2_3}, the authors further considered a scenario with a single SN and multiple channels, proposing three fundamental methods, i.e., resource ordering, routing, and distribution design for resource service time, to optimize the AoI-delay tradeoff. In \cite{shan_window}, the authors discussed the scenario with multiple SNs and one single channel, proposing a first come first serve (FCFS) method to serve the MUs, achieving a near-optimal AoI-delay tradeoff. It is important to note that while the existing literature on the AoI-delay tradeoff in MEC networks has made significant progress, the continuous-time models adopted in these works may not be suitable for practical MEC systems that operate in discrete time. Additionally, there is currently no research specifically addressing the scheduling problem for the scenario with multiple SNs and multiple channels, which is a most general scenario in practical applications.

\subsection{Main Contributions}
This paper focuses on a general discrete-time MEC network that encompasses one BS and multiple nearby SNs and MUs. The BS is responsible for scheduling multiple (time-division) channels to fetch time-sensitive messages from the SNs via uplinks or to serve the MUs by transmitting the requested messages from its local cache via downlinks. In this context, there exists a tradeoff between the AoI and the service delay, since the uplinks and downlinks in this MEC network share the same group of channels, which creates a competitive relationship between the two performance metrics. The main objective of this paper is to investigate this intricate tradeoff and develop a scheduling policy that achieves an optimal balance between AoI and service delay. The main contributions of this paper are summarized as follows:
\begin{itemize}
	\item We formulate the joint AoI and delay optimization for the MEC network as a sequential decision-making problem, whose design variables are high-dimensional, discrete-valued, and linearly constrained. However, to determine the achievable region or to validate the solution existence for this problem is NP-complete. To address this challenge, we characterize the superset and subset of the achievable region: First, we utilize the rate stability theorem to derive a superset of this region; then, we develop a novel stochastic policy to obtain a subset of this region, which is validated to be tight and possess the same set volume as the achievable region under specific conditions; finally, we leverage this subset to establish a sufficient condition for the solution existence of the considered problem.
	\item We propose an innovative Lyapunov drift optimization method to efficiently solve the formulated sequential decision-making problem, which is challenging due to the non-linear property of the objective function. First, we construct one linear (first-order) and one quadratic (second-order) Lyapunov functions for the AoI and the request queues, respectively. Then, we calculate the Lyapunov drifts for these functions and use properly designed weights to combine the two drifts with a penalty term, resulting in a mixed-order drift-plus-penalty formula. Finally, we employ dynamic programming (DP) to optimize this formula and derive the schedule decisions. Furthermore, by adopting the Lyapunov optimization technique, we provide theoretical evidence that the average AoI and average service delay achieved by our algorithm exhibit an $O(1/V)$ versus $O(V)$ tradeoff.
\end{itemize}
\begin{figure*}[t]
\centering
\includegraphics[width=6.9in]{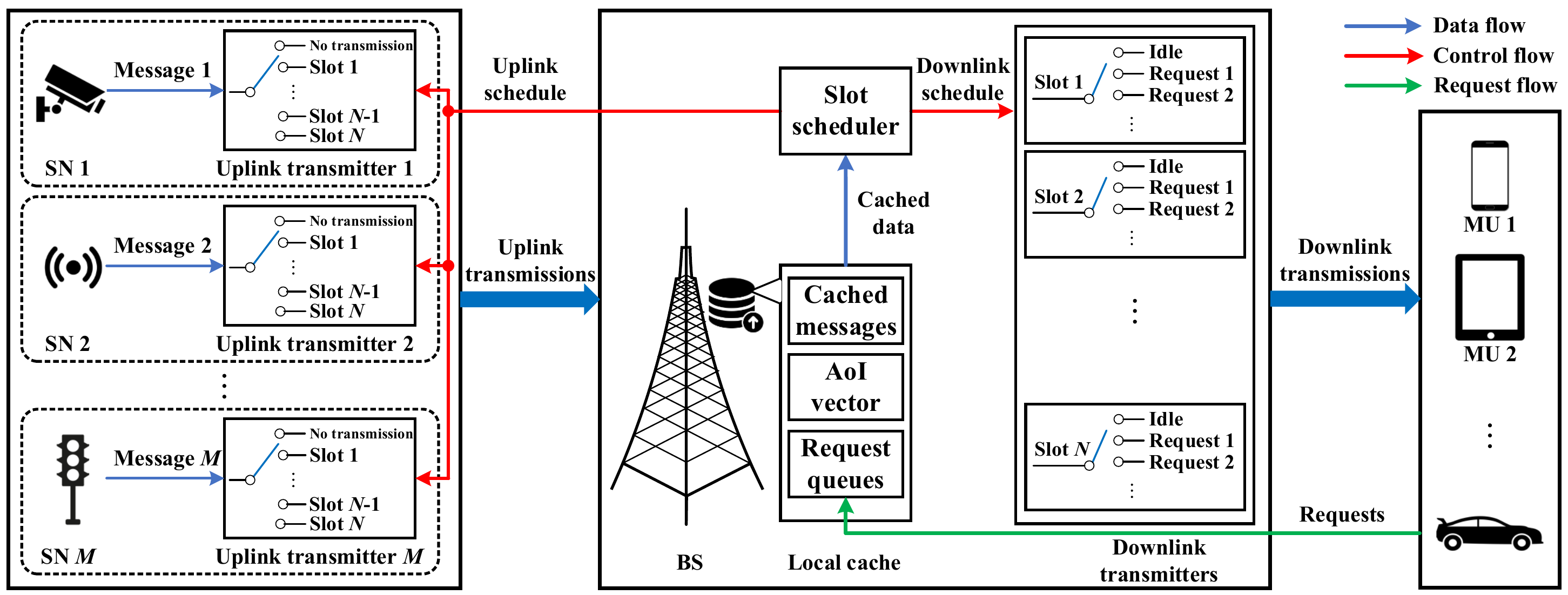}
\caption{System model for mobile edge caching network.}\label{fig:system_model}
\end{figure*}

The remainder of this paper is organized as follows. In Section \ref{sec2}, we present the system model and formulate the scheduling problem. Section \ref{sec3} analyzes the achievable region of the problem. In Section \ref{sec4}, we propose the mixed-order drift-plus-penalty algorithm and present theoretical evaluations of its performance. In Section \ref{sec6}, we present simulational evaluations of the proposed algorithm. Finally, Section \ref{sec7} concludes this paper.

\section{System Model and Problem Formulation}\label{sec2}

This section introduces the system model, including the transmission and information update models. Then, we define the average AoI and average service delay for the considered MEC system and formulate the corresponding scheduling problem.
\subsection{System Model}
Consider a MEC network, as depicted in Fig.~\ref{fig:system_model}, which consists of $M$ SNs, one BS, and multiple MUs. The SNs are located at $M$ different positions to monitor specific events and continuously pack the up-to-date monitored contents into $M$ messages, denoted as $\mathcal{M}=\{1,2,\dots,M\}$. The BS collects these messages from the SNs through wireless uplinks, stores them at its local cache, and tracks their ages. The MUs randomly send requests to the BS for downloading some of the messages in $\mathcal{M}$, and the BS queues these requests and selectively serves them by transmitting the corresponding cached messages through wireless downlinks. 

\begin{figure*}
\centering
\includegraphics[width=5.6in]{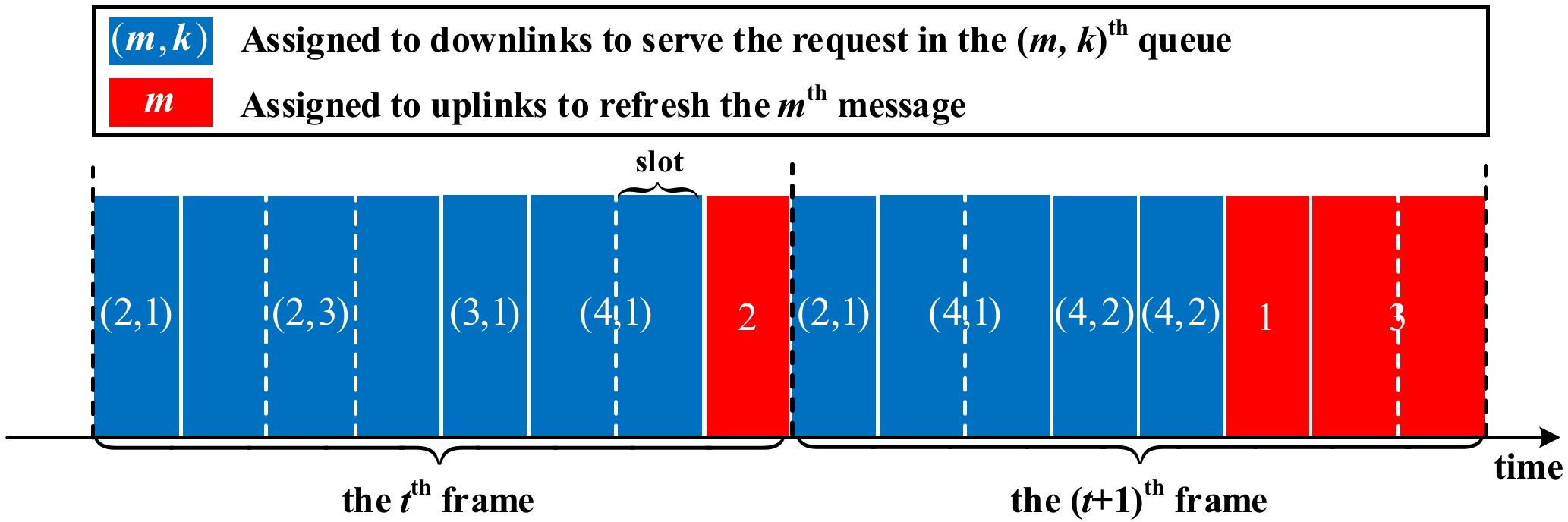}
\caption{An example of frame-based MEC system. The uplink transmissions of the 1\ts{st}, 2\ts{ed}, and 3\ts{rd} messages require 1, 1, and 2 slots, respectively. The downlink transmissions to refresh the (2,1)\ts{th}, (2,3)\ts{th}, (3,1)\ts{th}, (4,1)\ts{th}, and (4,2)\ts{th} messages require 1, 3, 1, 2, and 1 slots, respectively.}\label{fig:queues5}
\end{figure*}

As illustrated in Fig.~\ref{fig:queues5}, the MEC operates on a frame-based mechanism, where the aforementioned processes of message uploading, request queueing, and message downloading occur at the beginning of each frame. Moreover, each frame is composed of $N$ consecutive slots{\footnote[1]{In LTE ${\text{\cite{lte}}}$, each frame consists of 20 slots ($N=20$); while in 5G NR ${\text{\cite{5gnr}}}$, the frame structure is flexible, and $N$ is set as $10\cdot2^{i}$ with $i=0,1,2,\cdots$.}}. Within each frame, each slot can be allocated for either an uplink transmission from the SN to the BS or a downlink transmission from the BS to the MU. It is important to note that the allocations of slots within each frame are determined by the slot scheduler of the BS at the beginning of each frame. Additionally, a single uplink or downlink transmission may span multiple slots within a frame.

\subsubsection{Transmission model}In this part, we present the details about the uplink and downlink transmissions in the MEC network and derive the number of slots required for the transmission of each message.

\textbf{Uplink transmission}: The uplink channels between the SNs and the BS are considered to be quasi-static over each frame and experience slow variations across adjacent frames \cite{whoknows}. Specifically, we denote the channel power gain of the uplink from the $m$\ts{th} SN to the BS within the $t$\ts{th} frame as $g_m^{\text{UL}}(t)$ and model it as a stationary process with the following transition probability:
$$\text{Pr}\{g_m^{\text{UL}}(t+1)=g'|g_m^{\text{UL}}(t)=g\}=\text{Pr}_{m,\text{UL}}\{g'|g\},\ \forall t\in \mathbb{Z}_{>0},$$
where $\mathbb{Z}_{>0}$ is the set of all positive integers, and $\text{Pr}_{m,\text{UL}}\{g'|g\}$ is a constant representing the probability for $g_m^{\text{UL}}(t)$ transiting from state $g$ to state $g'$. Then, the maximum transmission rate over this uplink at the $t$\ts{th} frame is given as $B\log\left(1+\frac{P_{\text{SN}}g_m^{\text{UL}}(t)}{N_{\text{BS}}}\right)\Delta T$ bits per slot, where $B$, $\log(\cdot)$, $P_{\text{SN}}$, $N_{\text{BS}}$, and $\Delta T$ represent the available bandwidth of the MEC network, the logarithm function, the maximum transmission power at the SN\footnote{In general scenarios like LTE and 5G NR \cite{lte,5gnr}, dynamic adjustment of transmission power among different slots or frames is possible. However, for our specific optimization objectives of maximizing message freshness and service promptness, it is evident that prioritizing the utilization of the maximum transmission power is the preferred approach.}, the noise power at the BS, and the duration of one slot, respectively. Let $L_m$ denote the length (in bits) of the $m$\ts{th} message. Then, the number of slots required to upload the $m$\ts{th} message over this uplink within the $t$\ts{th} frame is calculated as
\begin{align}
\kappa_m^{\text{UL}}(t)\triangleq\Bigg\lceil\frac{L_m}{B\log\left(1+\frac{P_{\text{SN}}g_m^{\text{UL}}(t)}{N_{\text{BS}}}\right)\Delta T}\Bigg\rceil,
\end{align}
where $\lceil\cdot\rceil$ is the ceiling function and returns the smallest integer greater than or equal to the given number. Finally, we define the maximum value of $\kappa_m^{\text{UL}}(t)$ over $m$ and $t$ as $\hat{K}$, i.e., $\hat{K}\triangleq\max_{m\in\mathcal{M},t\in\mathbb{Z}_{>0}}\kappa_m^{\text{UL}}(t)$.

\textbf{Downlink transmission}: Similar to the uplink channels, the downlink channels between the BS and the MUs are also considered to be quasi-static over each frame and experience slow variations across adjacent frames. We denote the channel power gain of the downlink from the BS to the $u$\ts{th} MU within the $t$\ts{th} frame as $g_u^{\text{DL}}(t)$. Then, the maximum transmission rate over this downlink at the $t$\ts{th} frame is calculated as $B\log\Big(1+$ $\frac{P_{\text{BS}}g_u^{\text{DL}}(t)}{N_u}\Big)\Delta T$ bits per slot, where $P_{\text{BS}}$ and $N_u$ represent the maximum transmission power at the BS and the noise power at the $u$\ts{th} MU, respectively. Hence, the number of slots required for sending the $m$\ts{th} message over this downlink within the $t$\ts{th} frame is given as
\begin{align}
\kappa_{m,u}^{\text{DL}}(t)\triangleq\Bigg\lceil\frac{L_{m}}{B\log\left(1+\frac{P_{\text{BS}}g_u^{\text{DL}}(t)}{N_u}\right)\Delta T}\Bigg\rceil.
\end{align}
It is important to note that the BS would serve each MU as quickly as possible. Therefore, we assume that the value of $g_u^{\text{DL}}(t)$ remains constant during the short period (typically spanning several frames) leading up to the $u$\ts{th} MU being served, so is $\kappa_{m,u}^{\text{DL}}(t)$. Moreover, we define the maximum value of $\kappa_{m,u}^{\text{DL}}(t)$ over $m$, $u$, and $t$ as $\bar{K}$, i.e., $\bar{K}\triangleq\max_{m\in\mathcal{M},u,t\in\mathbb{Z}_{>0}}\kappa_{m,u}^{\text{DL}}$(t).
\subsubsection{Information update model}We first introduce three types of information stored in the local cache of the BS, i.e., the messages uploaded from the SNs, their corresponding ages, and the requests sent from the MUs.
\begin{itemize}
	\item \textbf{Cached messages}: The BS caches $M$ messages that are most recently uploaded from the SNs;
	\item \textbf{AoI vector}: The AoI vector stores the ages of the $M$ cached messages. Specifically, the AoI of the $m$\ts{th} message cached in the BS at the beginning of the $t$\ts{th} frame is denoted as $x_m(t)\in\mathbb{Z}_{>0}$. Then, the AoI vector is defined as $\bm{x}(t)\triangleq[x_1(t),x_2(t),\cdots,x_M(t)]^T$;
	\item \textbf{Request queues}: For each request from the MUs, it may demand any one of the total $M$ messages and the corresponding downlink transmission may take a duration ranging from $1$ to $\bar{K}$ slots. Therefore, the BS employs $M\bar{K}$ request queues to store the requests from all MUs, where the requests demanding the $m$\ts{th} message and requiring $k$ slots for downlink transmission are stored in the $(m,k)$\ts{th} request queue. We denote the length of the $(m,k)$\ts{th} request queue at the $t$\ts{th} frame as $q_{m,k}(t)$ and represent these $M\bar{K}$ request queues with a matrix $\bm{Q}(t)\in\mathbb{Z}_{\geq0}^{M\times \bar{K}}$, where $[\bm{Q}(t)]_{m,k}\triangleq q_{m,k}(t)$, and $\mathbb{Z}_{\geq0}$ represents the set of all non-negative integers.
\end{itemize}

After the BS makes the slot allocation decision, the above three elements are updated accordingly. We denote the slot allocation decision at the $t$\ts{th} frame as $\bm{A}(t)\in\mathbb{Z}_{\geq0}^{M\times(\bar{K}+1)}$ with $a_{m,k}(t)\triangleq[\bm{A}(t)]_{m,k}$: For $1\leq k\leq \bar{K}$, $a_{m,k}(t)$ represents the number of requests in the $(m,k)$\ts{th} queue to be served over downlinks within the $t$\ts{th} frame; for $k=\bar{K}+1$, $a_{m,k}(t)$ takes value from $\{0,1\}$, with $a_{m,\bar{K}+1}(t)=1$ indicating that the up-to-date version of the $m$\ts{th} message is to be uploaded over uplink within the $t$\ts{th} frame and $a_{m,\bar{K}+1}(t)=0$ indicating that it is not to be uploaded. With this notation, the information update models for these three elements are described as follows.
\begin{itemize}
	\item \textbf{Update of cached messages}: We replace the cached messages with their most recently uploaded version;
	\item \textbf{Update of AoI vector}: If the $m$\ts{th} message is uploaded over uplink within the $t$\ts{th} frame, i.e., $a_{m,\bar{K}+1}(t)=1$, the AoI of the $m$\ts{th} message is set to 1; otherwise, it increases by one. In summary, we have
        \begin{align}
        x_m(t+1)=x_m(t)+1-a_{m,\bar{K}+1}(t)x_m(t).\label{eq:x_m}
        \end{align}
    \item \textbf{Update of request queues}: The update of the request queues depends on the number of arrival and departure requests. For the $(m,k)$\ts{th} queue, the number of departure requests in the $t$\ts{th} frame is equal to $a_{m,k}(t)$. Additionally, the number of arrival requests, denoted as $c_{m,k}(t)$, is modeled as an independent and identically distributed (i.i.d.) random variable across $t$. We denote its mean value as $\lambda_{m,k}$ and its probability mass function (pmf) as $f_{m,k}$. Then, the update rule for $q_{m,k}(t)$ is given as
\begin{align}\label{eq:q_m_k}
q_{m,k}(t+1)\!=\!\max\{q_{m,k}(t)\!-\!a_{m,k}(t),0\}\!+\!c_{m,k}(t).
\end{align}
\end{itemize}

Remarkably, the slot allocation decision mentioned above are subject to the following constraints: First, the number of served requests in each request queue cannot exceed the number of its stored requests, i.e.,
\begin{align}
a_{m,k}(t)\leq\!q_{m,k}(t),\label{con:a_m_k_2}
\end{align}
for all $m\in\mathcal{M}$, $k\in\bar{\mathcal{K}}$, and $t\in\mathbb{Z}_{>0}$, where $\bar{\mathcal{K}}$ is defined as $\bar{\mathcal{K}}\triangleq\{1,2,\cdots,\bar{K}\}$; second, the slot allocation decision involves allocating $\sum_{m=1}^M\kappa_m^{\text{UL}}(t)a_{m,\bar{K}+1}(t)$ slots for uplink transmissions and $\sum_{m=1}^M\sum_{k=1}^{\bar{K}}ka_{m,k}(t)$ slots for downlink transmissions. Therefore, the total number of slots allocated in one frame should not exceed the available slots, i.e.,
\begin{align}\label{con:a_m_k_1}
	\sum_{m=1}^M\kappa_m^{\text{UL}}(t)a_{m,\bar{K}+1}(t)+\sum_{m=1}^M\sum_{k=1}^{\bar{K}}ka_{m,k}(t)\leq N.
\end{align}

\subsection{Problem Formulation}
This work aims to jointly optimize the average AoI and the average service delay of the randomly arrival requests, which are rigorously defined as follows.

\textbf{Average AoI}: To serve each request in the $(m,k)$\ts{th} queue, the BS first picks the $m$\ts{th} message in the local cache and then transmits it to the corresponding MU over downlink. Obviously, the AoI of this served request can be computed as the AoI of the $m$\ts{th} message stored in the cache plus one more frame required for the corresponding downlink transmission\footnote{We assume that the minimum unit of AoI and service delay is frame, instead of slot. Therefore, the transmission time is approximately considered to be one frame.}, i.e., $x_m(t)+1$ frames. Meanwhile, according to the slot allocation decision $\bm{A}(t)$, BS would serve $a_{m,k}(t)$ requests in the $(m,k)$\ts{th} request queue at the $t$\ts{th} frame. Hence, the sum AoI of the requests stored in the $(m,k)$\ts{th} queue and served at the $t$\ts{th} frame is calculated as $a_{m,k}(t)(x_m(t)+1)$. Then, considering the long-term average, the average AoI of all requests stored in $M\bar{K}$ request queues is calculated as
\begin{align*}
&\lim_{T\rightarrow\infty}\frac{\sum_{t=1}^T\sum_{m=1}^M\sum_{k=1}^{\bar{K}}a_{m,k}(t)(x_m(t)+1)}{\sum_{t=1}^T\sum_{m=1}^M\sum_{k=1}^{\bar{K}}c_{m,k}(t)}\\
=&\frac{1}{\sum_{m=1}^M\!\sum_{k=1}^{\bar{K}}\!\lambda_{m,k}}\!\lim_{T\rightarrow\infty}\!\frac{1}{T}\sum_{t=1}^T\!\sum_{m=1}^M\!\sum_{k=1}^{\bar{K}}a_{m,k}(t)(x_m(t)+1),
\end{align*}
where we use the fact $\lim_{T\rightarrow\infty}\frac{1}{T}c_{m,k}(t)=\lambda_{m,k}$, $\sum_{t=1}^T\sum_{m=1}^M\sum_{k=1}^{\bar{K}}$ $a_{m,k}(t)(x_m(t)+1)$ represents the overall AoI of the arrival requests, and $\sum_{t=1}^T\sum_{m=1}^M\sum_{k=1}^{\bar{K}}c_{m,k}(t)$ is the total number of the arrival requests.

\textbf{Average service delay}: The service delay for each request is the sum of the queueing delay and the downlink transmission delay. By the queueing theorem \cite{neely_queue}, the average queueing delay for the requests stored in the $(m,k)$\ts{th} queue is equal to the average queue length, which is given as $\lim_{T\rightarrow\infty}\frac{1}{T}\sum_{t=1}^Tq_{m,k}(t)$. Additionally, the average downlink transmission delay is fixed as one frame. Thus, the average service delay for the requests stored in the $(m,k)$\ts{th} queue is $\lim_{T\rightarrow\infty}\frac{1}{T}\sum_{t=1}^Tq_{m,k}(t)+1$. Considering the long-term average, the average service delay for all requests stored in $M\bar{K}$ request queues is calculated as
\begin{align*}
&\lim_{T\rightarrow\infty}\frac{\sum_{m=1}^M\sum_{k=1}^{\bar{K}}\lambda_{m,k}T(\lim_{T\rightarrow\infty}\frac{1}{T}\sum_{t=1}^Tq_{m,k}(t)+1)}{\sum_{t=1}^T\sum_{m=1}^M\sum_{k=1}^{\bar{K}}c_{m,k}(t)}\\
=&\frac{1}{\sum_{m=1}^M\sum_{k=1}^{\bar{K}}\lambda_{m,k}}\lim_{T\rightarrow\infty}\frac{1}{T}\sum_{t=1}^T\sum_{m=1}^M\sum_{k=1}^{\bar{K}}\lambda_{m,k}q_{m,k}(t)+1,
\end{align*}
where $\sum_{m=1}^M\sum_{k=1}^{\bar{K}}\lambda_{m,k}T(\lim_{T\rightarrow\infty}\frac{1}{T}\sum_{t=1}^Tq_{m,k}(t)+1)$ represents the overall service delay of the arrival requests.

From the above analysis, we now formulate the optimization problem to jointly minimize both the average AoI and the average service delay as the following sequential decision-making problem:
\begin{align}
\begin{split}\label{newnew}
\textbf{(P1)}&\min_{\bm{A}(t)}\ \lim_{T\rightarrow\infty}\frac{1}{T}\sum_{t=1}^T\sum_{m=1}^M\sum_{k=1}^{\bar{K}}\Big(Va_{m,k}(t)(x_m(t)\!+\!1)\!\\
&\qquad\ +\!\lambda_{m,k}q_{m,k}(t)\Big)
\end{split}\\
&\ \ \text{s.t.}\ \ \eqref{eq:x_m},\eqref{eq:q_m_k},\eqref{con:a_m_k_2},\eqref{con:a_m_k_1},\nonumber
\end{align}
where, $V\in\mathbb{R}_{\geq0}$, with $\mathbb{R}_{\geq0}$ being the set of all non-negative real numbers, is a tradeoff parameter to balance AoI and service delay. Notably, according to \cite{puterman}, any solution to problem {\bf (P1)} can be characterized by a slot allocation policy $\pi$, which determines the value of $\bm{A}(t)$ based on the historical information $h(t)\triangleq(\bm{x}(1),\bm{Q}(1),\bm{A}(1),$ $\bm{x}(2),\bm{Q}(2),\bm{A}(2),\cdots,\bm{x}(t),$ $\bm{Q}(t))$ and takes the form of $\pi:h(t)\rightarrow\bm{A}(t)$. Therefore, solving problem {\bf (P1)} is equivalent to finding a feasible slot allocation policy that minimizes \eqref{newnew}. Here, a slot allocation policy is considered feasible if the value of $\bm{A}(t)$ under this policy satisfies the constraints in \eqref{eq:x_m}, \eqref{eq:q_m_k}, \eqref{con:a_m_k_2}, and \eqref{con:a_m_k_1} for all $t\in\mathbb{Z}_{t>0}$.

\begin{Rem}\label{rem21}
The linear constraints in \eqref{con:a_m_k_1} makes it difficult to determine the existence of the solution to problem {\bf (P1)}, which is actually an NP-complete problem {\normalfont\cite{bertsekas_dp}}. Additionally, coupled with the linear constraints, the high-dimensional and discrete-valued nature of design variable $\bm{A}(t)$ makes problem {\bf (P1)} challenging to be solved and the existing tools cannot efficiently address these challenges:
\begin{itemize}
	\item Dynamic programming suffers from the curse of dimensionality and cannot handle the problems with high-dimensional and discrete-valued design variables {\normalfont \cite{bertsekas_dp}}; 
	\item Deep reinforcement learning (DRL) cannot efficiently solve problem {\bf (P1)} since the linear constraints in \eqref{con:a_m_k_1} strongly limit the feasible actions to only $1/(M(\bar{K}+1))!$ of all the possible ones and make the convergence of DRL during the offline training phase extremely difficult {\normalfont \cite{liran1,liran2}};
	\item Conventional Lyapunov drift optimization method also cannot be directly applied to solve problem {\bf (P1)}. Specifically, the average AoI term in the objective function \eqref{newnew} is the product of the design variable and a linear function of AoI, which does not fit the linear form that conventional Lyapunov methods are designed to handle {\normalfont \cite[Theorem 4.2]{neely_queue}}.
\end{itemize}
\end{Rem}

\begin{figure}
\centering
\includegraphics[width=3.4in]{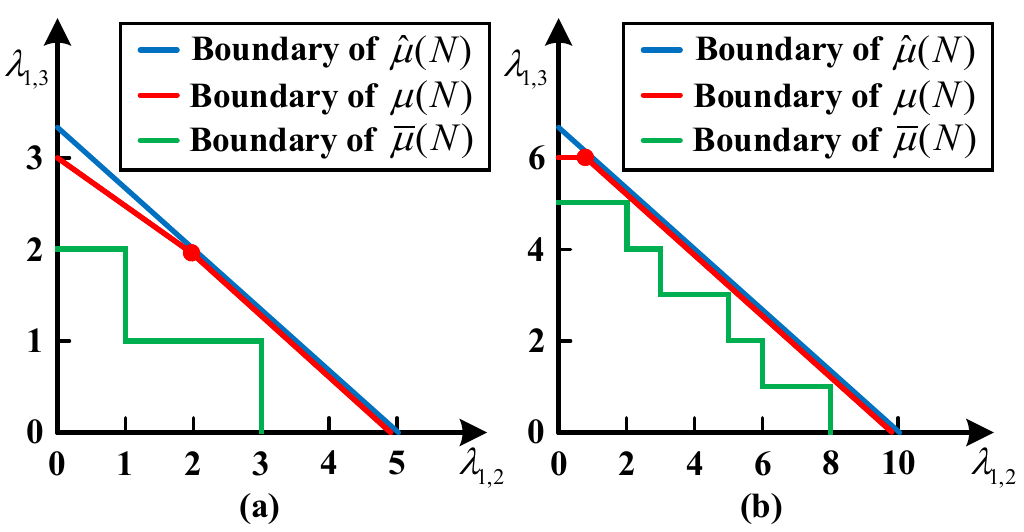}
\caption{Boundaries of the achievable region $\mu(N)$, its superset $\hat{\mu}(N)$, and its subset $\bar{\mu}(N)$ with $M=1$, $\hat{K}=1$, $\bar{K}=3$, and $\lambda_{1,1}=0$. (a) MEC network with $N=10$; (b) MEC network with $N=20$.}\label{fig:boundary}	
\end{figure}

\section{Achievable Region Analyses}\label{sec3}
In this section, we first define the achievable region of problem {\bf (P1)}. Then, we characterize this region by studying both its superset and subset. Finally, we develop a sufficient condition to determine whether a solution to problem {\bf (P1)} exists.

\subsubsection{Achievable region of problem \rm{\textbf{(P1)}}}
Denote $\boldsymbol{\lambda}\in\mathbb{R}_{\geq0}^{M\times\bar{K}}$ as the mean arrival matrix with $[\boldsymbol{\lambda}]_{(m,k)}\triangleq\lambda_{m,k}$. Next, with fixed $\boldsymbol{\lambda}$ and $N$, we denote the value of problem \rm{\textbf{(P1)}} under a general slot allocation policy $\pi$ as $f_{\pi}(\boldsymbol{\lambda},N)$. Then, we define the achievable region of problem \rm{\textbf{(P1)}} as follows.
\begin{Def}\label{defonly}
The mean arrival matrix $\boldsymbol{\lambda}$ is achievable if there exists a feasible slot allocation policy $\pi$ such that $f_{\pi}(\boldsymbol{\lambda},N)<\infty$ holds. Then, the achievable region of problem {\rm\textbf{(P1)}}, denoted by $\mu(N)$, is defined as the set containing all achievable $\boldsymbol{\lambda}$, i.e.,
\begin{align}\label{def:lambda_N}
\mu(N)\triangleq\Big\{\boldsymbol{\lambda}\Big|\min_{\pi}f_{\pi}(\boldsymbol{\lambda},N)<\infty, \boldsymbol{\lambda}\succeq 0\Big\}.
\end{align} 
\end{Def}
\begin{Rem}
Based on Definition \ref{defonly}, a solution to problem {\rm{\textbf{(P1)}}} exists only if the mean arrival matrix $\boldsymbol{\lambda}$ is achievable. However, validating whether $\boldsymbol{\lambda}$ is achievable can be extremely difficult, as it involves checking the value of $f_{\pi}(\boldsymbol{\lambda},N)$ for all feasible slot allocation policies. An alternative approach is to first determine the achievable region of problem {\rm{\textbf{(P1)}}} and then check whether $\boldsymbol{\lambda}$ lies within this region. However, it can be validated that determining the achievable region of problem {\rm{\textbf{(P1)}}} is also an NP-complete task {\normalfont\cite{bertsekas_dp}}, compelling us to resort to characterizing this region.
\end{Rem}

\subsubsection{Achievable region characterization} To characterize the achievable region of problem {\rm{\textbf{(P1)}}}, we derive its superset and subset, and then analyze their properties.

First, we utilize the rate stability theorem \cite[Theorem 2.4]{neely_queue} to derive a superset of $\mu(N)$. According to this theorem, we have the following results: (1) The arrival rate of the $(m,k)$\ts{th} request queue in problem {\bf (P1)} is equal to $k\lambda_{m,k}$, where $k$ is the number of required slots to serve one request in this queue and $\lambda_{m,k}$ is the arrival mean of this queue; (2) the maximum allowable departure rate of $M\bar{K}$ request queues is equal to the number of slots in one frame, i.e., $N$; and (3) if a mean arrival matrix $\bm{\lambda}$ belongs to the achievable region $\mu(N)$, the total arrival rate of all $M\bar{K}$ request queues should not exceed the maximum allowable departure rate, i.e.,  $\sum_{m=1}^M\sum_{k=1}^{\bar{K}}k\lambda_{m,k}\leq N$. Based on these results, we define a set $\hat{\mu}(N)$ as
\begin{align}\label{def:hat_mu}
	\hat{\mu}(N)\triangleq\Big\{\boldsymbol{\lambda}\Big|\sum_{m=1}^M\sum_{k=1}^{\bar{K}}k\lambda_{m,k}\leq N, \boldsymbol{\lambda}\succeq 0\Big\}.
\end{align}
Apparently, $\hat{\mu}(N)$ serves as a superset of $\mu(N)$, i.e., $\mu(N)\subseteq\hat{\mu}(N)$. Note that the boundary of $\hat{\mu}(N)$ is a hyperplane characterized by the equality $\sum_{m=1}^M\sum_{k=1}^{\bar{K}}k\lambda_{m,k}=N$ (see the blue curves in Fig. \ref{fig:boundary}), indicating that the boundary of the achievable region $\mu(N)$ lies below or on this hyperplane. 

Then, we propose the following theorem to derive a subset of $\mu(N)$.
\begin{Thm}\label{thm1.1}
Define a set $\bar{\mu}(N)$ as 
\begin{align}\label{def:mu_pi_s}
\bar{\mu}(N)\triangleq\Big\{\bm{\lambda}\Big|\hat{K}+\sum_{k=1}^{\bar{K}}k\lceil\lambda_k\rceil\leq N, \boldsymbol{\lambda}\succeq 0\Big\},
\end{align}
with $\lambda_k\triangleq\sum_{m=1}^M\lambda_{m,k}$. Then, $\bar{\mu}(N)$ is a subset of $\mu(N)$, i.e., $\bar{\mu}(N)\subseteq\mu(N)$. Moreover, the set volumes of $\bar{\mu}(N)$ and $\mu(N)$, denoted as {\normalfont $\text{Vol}(\bar{\mu}(N))$} and {\normalfont $\text{Vol}(\mu(N))$}, satisfies
\begin{align*}
\normalfont \lim_{N\rightarrow\infty}\frac{\text{Vol}(\bar{\mu}(N))}{\text{Vol}(\mu(N))}=1,
\end{align*}
with $\normalfont \text{Vol}(\bar{\mu}(N))\!\triangleq\!\int_{\boldsymbol{\lambda}\in\bar{\mu}(N)}\text{d}\boldsymbol{\lambda}$ and $\normalfont\text{Vol}(\mu(N))\!\triangleq\!\int_{\boldsymbol{\lambda}\in\mu(N)}\text{d}\boldsymbol{\lambda}$.
\end{Thm}
\begin{IEEEproof}[Sketch of proof]
To prove Theorem \ref{thm1.1}, we first propose a stochastic slot allocation policy $\pi_s(\bm{\lambda}):\bm{A}(t)\rightarrow[0,1]$, which allocates the first $(\hat{K}+\sum_{k=1}^{\bar{K}}k\lceil\lambda_k\rceil)$ slots in each frame for the uplink and downlink transmissions. Then, we prove that for any $\bm{\lambda}\in\bar{\mu}(N)$, $f_{\pi_s(\bm{\lambda})}(\bm{\lambda},N)<\infty$ holds, which implies $\bar{\mu}(N)\subseteq\mu(N)$. Finally, based on the definitions of $\bar{\mu}(N)$ and $\mu(N)$, we prove $\lim_{N\rightarrow\infty}\text{Vol}(\bar{\mu}(N))/\text{Vol}(\mu(N))=1$. Please check Appendix \ref{app:A} for more details.
\end{IEEEproof}

Based on Theorem \ref{thm1.1}, $\bar{\mu}(N)$ serves as a subset of $\mu(N)$, and as $N$ increases, the set volume of $\bar{\mu}(N)$ asymptotically approaches that of $\mu(N)$. Moreover, the boundary of subset $\bar{\mu}(N)$ is characterized by the equality $\hat{K}+\sum_{k=1}^{\bar{K}}k\lceil\lambda_k\rceil=N$ and thus has a piecewise linear shape (see the green curves in Fig. \ref{fig:boundary}), indicating that the boundary of $\mu(N)$ lies above or on this piecewise linear surface.

In summary, the achievable region of problem {\bf (P1)}, i.e., $\mu(N)$, can be characterized by superset $\hat{\mu}(N)$ and subset $\bar{\mu}(N)$. Additionally, Theorem \ref{thm1.1} provides a sufficient condition for the solution existence of problem {\bf (P1)}: If the condition $\hat{K}+\sum_{k=1}^{\bar{K}}k\lceil\lambda_k\rceil\leq N$ is satisfied, then problem {\bf (P1)} has at least one solution, and this solution is represented by the policy $\pi_s(\bm{\lambda})$ as introduced in Appendix \ref{app:A}.

\section{Mixed-Order Drift-Plus-Penalty Algorithm}\label{sec4}
In this section, we first analyze the characteristics of the two terms in the objective function \eqref{newnew} of problem {\bf (P1)}. Next, leveraging these characteristics and the Lyapunov drift optimization \cite{neely_queue}, we introduce the linear and quadratic Lyapunov functions, along with a penalty term. Then, we combine the drifts of these Lyapunov functions with the penalty term to develop a mixed-order drift-plus-penalty algorithm. Finally, we conduct the performance analysis of the proposed algorithm.

\subsection{Lyapunov Functions and Drifts}
The objective function \eqref{newnew} of problem {\bf (P1)} contains an AoI term and a service delay term. Specifically, the service delay term, given by $\sum_{m=1}^M\sum_{k=1}^{\bar{K}}\lambda_{m,k}q_{m,k}(t)$, exhibits a linear relationship with respect to the request queues $\bm{Q}(t)$. Hence, by adopting Lyapunov drift optimization to this term, we construct a quadratic Lyapunov function with respect to $\bm{Q}(t)$ as \cite[Theorem 4.1]{neely_queue}
\begin{align}
&L(\bm{Q}(t))\triangleq\frac{1}{2}\sum_{m=1}^M\sum_{k=1}^{\bar{K}}q_{m,k}^2(t),\label{def:l_q}
\end{align}
and define the corresponding Lyapunov drift under a general slot allocation policy $\pi$ as
\begin{align}
\begin{split}\label{def:drift_q}
&\Delta_{\pi}(L(\bm{Q}(t)))\\
\triangleq&\mathbb{E}_{\pi,c_{m,k}(t)}\left[L(\bm{Q}(t+1))-L(\bm{Q}(t))|\bm{x}(t),\bm{Q}(t)\right].
\end{split}
\end{align}

However, the AoI term in the objective function \eqref{newnew} of problem {\bf (P1)}, given by $\sum_{m=1}^M\sum_{k=1}^{\bar{K}}$ $a_{m,k}(t)(x_m(t)$ $+1)$, is the product of the design variable $a_{m,k}(t)$ and a linear function of the AoI $x_m(t)$, and thus cannot be handled by conventional Lyapunov methods \cite[Theorem 4.2]{neely_queue}. To address this challenge, we first propose a linear Lyapunov function of $\bm{x}(t)$ and a penalty term. Specifically, the linear Lyapunov function is defined as
\begin{align}
&L(\bm{x}(t))\triangleq\sum_{m=1}^Mx_m(t),\label{def:l_x}
\end{align}
and the corresponding Lyapunov drift under a general slot allocation policy $\pi$ is defined as
\begin{align}
\begin{split}\label{def:drift_x}
&\Delta_{\pi}(L(\bm{x}(t)))\\
\triangleq&\mathbb{E}_{\pi,c_{m,k}(t)}\left[L(\bm{x}(t+1))-L(\bm{x}(t))\Big|\bm{x}(t),\bm{Q}(t)\right].
\end{split}
\end{align}
The proposed penalty term is defined as the conditional expectation of the AoI term under a general slot allocation policy $\pi$, i.e., $\sum_{m=1}^M\sum_{k=1}^{\bar{K}}\mathbb{E}_{\pi}\left[a_{m,k}(t)|\bm{x}(t),\bm{Q}(t)\right]$ $(x_m(t)+1)$. Next, we combine the linear drift $\Delta_{\pi}(L(\bm{x}(t)))$ in \eqref{def:drift_x}, the quadratic drift $\Delta_{\pi}(L(\bm{Q}(t)))$ in \eqref{def:drift_q}, and the penalty term to obtain the ``mixed-order drift-plus-penalty", i.e.,
\begin{align}
\begin{split}\label{eq:drift_pi1}
&\Delta_{\pi}(L(\bm{Q}(t)))+V\Big(V_0\Delta_{\pi}(L(\bm{x}(t)))\\
&+\sum_{m=1}^M\sum_{k=1}^{\bar{K}}\mathbb{E}_{\pi}\left[a_{m,k}(t)|\bm{x}(t),\bm{Q}(t)\right](x_m(t)+1)\Big),
\end{split}
\end{align}
where $V_0$ is a positive constant. Then, we find an upper bound for the mixed-order drift-plus-penalty with the following proposition.
\begin{Prop}\label{lemma1}
For any positive constant $V_0$, the defined mixed-order drift-plus-penalty in \eqref{eq:drift_pi1} is upper bounded by
\begin{align}
\begin{split}\label{eq:drift_pi2}
&C\!-\!\!\!\sum_{m=1}^M\!\sum_{k=1}^{\bar{K}}\!\!\lambda_{m,k}q_{m,k}(t)\!\left(\mathbb{E}_{\pi}\!\left[a_{m,k}(t)|\bm{x}(t),\bm{Q}(t)\right]\!-\!\lambda_{m,k}\right)\\
&-VV_0\sum_{m=1}^M\mathbb{E}_{\pi}[a_{m,\bar{K}+1}(t)|\bm{x}(t),\bm{Q}(t)]x_m(t)+VV_0M\\
&+V\sum_{m=1}^M\sum_{k=1}^{\bar{K}}\mathbb{E}_{\pi}\left[a_{m,k}(t)|\bm{x}(t),\bm{Q}(t)\right](x_m(t)+1),
\end{split}
\end{align}
where $C$ is given as 
\begin{align*}
C\!\triangleq\!\frac{1}{2}\!\sum_{m=1}^M\!\sum_{k=1}^{\bar{K}}\!\lambda_{m,k}\mathbb{E}_{c_{m,k}(t)}[c_{m,k}^2(t)]\!+\!\frac{1}{2}\max_{m\in\mathcal{M},k\in\bar{\mathcal{K}}}\!\lambda_{m,k}\!\Big\lceil\frac{N}{k}\Big\rceil^2\!\!.
\end{align*}
\end{Prop}
\begin{IEEEproof}
Please see Appendix \ref{app:B}.	
\end{IEEEproof}

Remarkably, the upper bound \eqref{eq:drift_pi2} proposed in Proposition \ref{lemma1} now serves as the new objective function in problem {\bf (P1)}, replacing the original objective function \eqref{newnew}. We will show that by minimizing this upper bound, we can effectively control both the average AoI and the average service delay in problem {\bf (P1)}, thereby overcoming the non-linearity challenge posed by the AoI term in the objective function \eqref{newnew} of problem {\bf (P1)}.

\subsection{Algorithm}
Our proposed mixed-order drift-plus-penalty algorithm follows the same principle as conventional Lyapunov methods \cite{neely_queue} in making slot allocation decisions in each frame, which consists of two steps: First, it calculates the values of $\bm{x}(t)$ and $\bm{Q}(t)$ based on their update rules in \eqref{eq:x_m} and \eqref{eq:q_m_k}; then, it obtains the slot allocation decision $\bm{A}(t)$ that minimizes the upper bound of the mixed-order drift-plus-penalty in \eqref{eq:drift_pi2} and simultaneously satisfies the constraints in \eqref{con:a_m_k_2} and \eqref{con:a_m_k_1}. In other words, the proposed algorithm obtains the value of $\bm{A}(t)$ by solving the following problem.
\begin{align}
\textbf{(P2)}\ &\argmin_{\bm{A}(t)}\ \ \eqref{eq:drift_pi2},\nonumber\\
&\quad\ \text{s.t.}\quad\ \ \eqref{con:a_m_k_2},\eqref{con:a_m_k_1},\nonumber
\end{align}
which can be reframed as
\begin{align*}
\begin{split}
\textbf{(P3)}\ &\argmax_{\bm{A}(t)}\ \sum_{m=1}^M\!\sum_{k=1}^{\bar{K}}[\lambda_{m,k}q_{m,k}(t)-V\!(x_m(t)+\!1)]a_{m,k}(t)\\
&\qquad\qquad +\!VV_0\sum_{m=1}^Mx_m(t)a_{m,\bar{K}+1}(t)
\end{split}\\
&\quad\ \text{s.t.}\quad\ \ \eqref{con:a_m_k_2},\eqref{con:a_m_k_1}.\nonumber
\end{align*}

In problem {\bf{(P3)}}, variables $a_{m,k}(t)$, $k\!\in\bar{\mathcal{K}}$, are bounded due to the constraints in \eqref{con:a_m_k_2}, and $a_{m,\bar{K}+1}(t)$ takes value from set $\{0,1\}$. Thus, problem {\bf{(P3)}} is a mixture of the bounded knapsack problem and the 0-1 knapsack problem \cite{kp}, and can be efficiently solved using DP algorithm within pseudo-polynomial time. The specific algorithm can be found in \cite{kp} and is omitted in this paper.

Finally, we summarize the mixed-order drift-plus-penalty algorithm in Algorithm \ref{alg:dpp}, where $T_0$ denotes the end scheduling frame, the values of $\bm{x}(t)$ and $\bm{Q}(t)$ are derived in lines 1 and 6, and the slot allocation decisions $\bm{A}(t)$ are determined in line 3. Notably, in Algorithm \ref{alg:dpp}, the value of the slot allocation decision $\bm{A}(t)$ solely depends on the values of $\bm{x}(t)$ and $\bm{Q}(t)$. As a result, we can represent the corresponding slot allocation policy under Algorithm \ref{alg:dpp} as $\pi_m: \bm{x}(t) \times \bm{Q}(t) \rightarrow \bm{A}(t)$, which is derived by mapping the values of $\bm{x}(t)$ and $\bm{Q}(t)$ to the corresponding solution of problem {\bf (P3)}.

\begin{algorithm}[t]
\caption{Proposed mixed-order drift-plus-penalty algorithm to solve problem {\bf (P1)}}\label{alg:dpp}
\begin{algorithmic}[1]
\STATE Initialize $\bm{x}(1)$ and $\bm{Q}(1)$ as $0^{M\times 1}$ and $0^{M\!\times\bar{K}}$, respectively.
\STATE \textbf{for} $t=1,2,\cdots, T_0$
\STATE \quad Based on the values of $\bm{x}(t)$ and $\bm{Q}(t)$, adopt DP
\STATEx \quad algorithm \cite{kp} to solve problem {\bf{(P3)}} and derive the
\STATEx \quad value of $\bm{A}(t)$;
\STATE \quad Execute the slot allocation decision $\bm{A}(t)$ at the BS;
\STATE \quad Observe the values of $c_{m,k}(t)$ for all $m\!\in\!\mathcal{M}$ and $k\!\in\bar{\mathcal{K}}$
\STATEx \quad at the BS;
\STATE \quad Derive the values of $\bm{x}(t\!+\!1)$ and $\bm{Q}(t\!+\!1)$ based on \eqref{eq:x_m}, 
\STATEx \quad \eqref{eq:q_m_k}, and the values of the observed $c_{m,k}(t)$;
\STATE \textbf{end for}
\end{algorithmic}
\end{algorithm}

\subsection{Performance Analysis}\label{sec5}
To evaluate the performance of the proposed Algorithm \ref{alg:dpp}, we first derive an upper bound on the expected value of \eqref{eq:drift_pi1} under this algorithm.
\begin{Prop}\label{prop5}
For any $\bm{\lambda}\in\bar{\mu}(N)$, we denote $\epsilon(\bm{\lambda})\in\mathbb{R}_{\geq0}$ as the maximum value satisfying $\bm{\lambda}+\epsilon(\bm{\lambda})\cdot1^{M\times {\bar{K}}}\in\bar{\mu}(N)$\footnote{Based on the definition of $\bar{\mu}(N)$, the value of $\epsilon(\bm{\lambda})$ is the solution to $\hat{K}+\sum_{k=1}^{\bar{K}}k\lceil\lambda_k\rceil+\frac{\bar{K}(\bar{K}+1)}{2}\epsilon(\bm{\lambda})=N$ and thus can be derived by bisection search algorithm.}. Then, for any $\epsilon\in[0,\epsilon(\bm{\lambda})]$, we have
\begin{align}
&\mathbb{E}_{\pi_m}[\eqref{eq:drift_pi1}|_{\pi=\pi_m}]\nonumber\\
\begin{split}\label{eq:drift_pi_d3}
\!\!\leq&C+V\Big(V_0M+\sum_{m=1}^M\sum_{k=1}^{\bar{K}}(\lambda_{m,k}+\epsilon)\Big)\\
&-\epsilon\sum_{m=1}^M\sum_{k=1}^{\bar{K}}\lambda_{m,k}\mathbb{E}_{\bm{Q}(t)\sim\pi_m(\bm{Q}(t))}[q_{m,k}(t)]\\
&-V\!\!\sum_{m=1}^M\!\sum_{k=1}^{\bar{K}}\!\left(\!\!\frac{V_0}{M\bar{K}}\!-\!(\!\lambda_{m,k}\!+\!\epsilon)\!\!\right)\!\mathbb{E}_{\bm{x}(t)\sim\pi_m(\bm{x}(t))}\![x_m(t)].
\end{split}
\end{align}
Here, $\mathbb{E}_{\pi_m}[\eqref{eq:drift_pi1}|_{\pi=\pi_m}]$ represents the expected value of $\eqref{eq:drift_pi1}$ when the slot allocation policy $\pi_m$ is adopted, $\pi_m(\bm{x}(t))$ and $\pi_m(\bm{Q}(t))$ represent the distributions of $\bm{x}(t)$ and $\bm{Q}(t)$ under policy $\pi_m$, respectively. 
\end{Prop}

\begin{IEEEproof}
Please see Appendix \ref{app:B2}.
\end{IEEEproof}

Then, we use the derived upper bound in Proposition \ref{prop5} to evaluate the performance of the proposed Algorithm \ref{alg:dpp}. The results are concluded in the following theorem.
\begin{Thm}\label{thm5.1}
Under the proposed {\normalfont Algorithm \ref{alg:dpp}}, the average AoI is upper bounded by
\begin{align}\label{eq:aoi_final}
\begin{split}
&\frac{1}{\sum_{m=1}^M\sum_{k=1}^{\bar{K}}\lambda_{m,k}}\Bigg(\max_{m\in\mathcal{M},k\in\bar{\mathcal{K}}}\left(\lambda_{m,k}+\epsilon(\bm{\lambda}\right))M^2{\bar{K}}\\
&+\sum_{m=1}^M\sum_{k=1}^{\bar{K}}\lambda_{m,k}+\frac{C}{V}\Bigg),
\end{split}
\end{align}
and the average service delay is upper bounded by
\begin{align}
\begin{split}\label{eq:delay_final}
&\frac{1}{\epsilon(\bm{\lambda})\!\sum_{m=1}^M\!\sum_{k=1}^{\bar{K}}\!\lambda_{m,k}}\Bigg(\!\!\Big(\!\max_{m\in\mathcal{M},k\in\bar{\mathcal{K}}}\!\left(\lambda_{m,k}\!+\!\epsilon(\bm{\lambda}\right))M^2{\bar{K}}\!\\
&+\!\sum_{m=1}^M\sum_{k=1}^{\bar{K}}(\lambda_{m,k}+\epsilon(\bm{\lambda}))\Big)V+\!C\!\Bigg)+1.
\end{split}
\end{align}
\end{Thm}
\begin{IEEEproof}
Please see Appendix \ref{app:C}.
\end{IEEEproof}
Based on \eqref{eq:aoi_final} and \eqref{eq:delay_final}, both the average AoI and the average service delay are upper bounded, and in general scenarios where the mean arrival matrix $\bm{\lambda}$ is fixed, the values of their upper bounds are $O(1/V)$ and $O(V)$, respectively.

\section{Simulation Results}\label{sec6}
\subsection{Simulation Setup}
In this section, we evaluate the performance of the proposed mixed-order drift-plus-penalty algorithm by comparing it with three widely-adopted algorithms:
\begin{itemize}
	\item Fixed window algorithm \cite{shan_window}: the state-of-the-art near-optimal policy for the case of $N=1$. It uploads messages via uplinks as soon as their ages reach certain thresholds and serves requests via downlinks using the FCFS mechanism. We extend this algorithm to cases with $N>1$ by introducing a naive parallel mechanism to schedule the $N$ slots within a frame; 
	\item DRL \cite{liran1,liran2}: a near-optimal policy for scenarios with small $N$ values. However, its efficiency decreases significantly when $N$ becomes large, due to the substantial increase in the size of neural networks used by the actor and critic of DRL;
	\item $\pi_s(\bm{\lambda})$: the benchmark algorithm proposed in Appendix \ref{app:A}.
\end{itemize}

We set the simulation parameters as follows: the value of $\kappa_m^{\text{UL}}(t)$ is set to either 1 or 2 with equal probability, the pmf $f_{m,k}$ follows a Poisson distribution, with its mean value $\lambda_{m,k}$ randomly sampled from a uniform distribution over the interval $[0,1]$.

\subsection{Performance Evaluashtion}
\begin{figure*}[htb]
\centering
\includegraphics[width=5.4in]{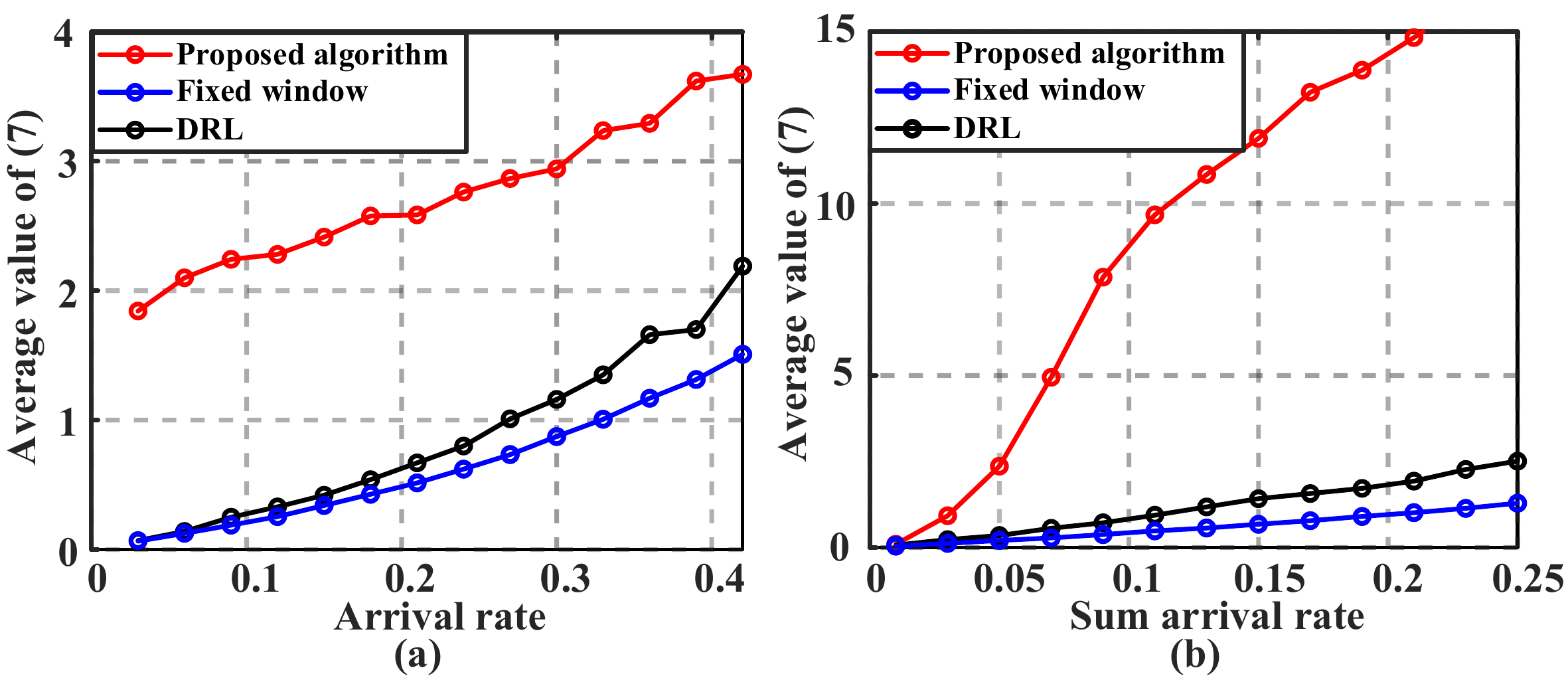}
\caption{Arrival rate vs. average value of \eqref{newnew} of various algorithms. (a) The scenario with $M=1$, $N=1$, $\bar{K}=1$; (b) The scenario with $M=10$, $N=1$, $\bar{K}=1$.}\label{fig1}
\end{figure*}
\begin{figure*}[htb]
\centering
\includegraphics[width=6.3in]{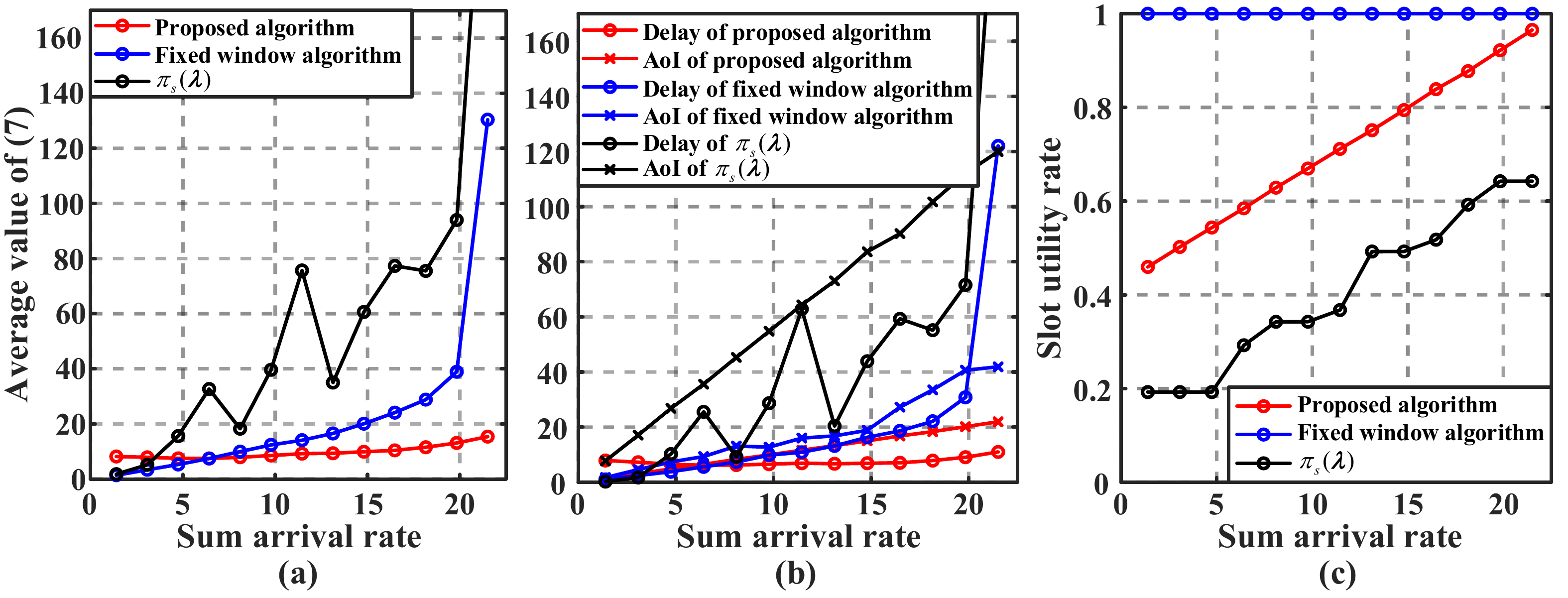}
\caption{Performances of various algorithms in the scenario with $M=10$, $N=40$, and $\bar{K}=3$. (a) Sum arrival rate vs. average value of \eqref{newnew}; (b) Sum arrival rate vs. delay/AoI; (c) Sum arrival rate vs. slot utility rate.}\label{fig2}
\end{figure*}
\begin{figure}[htb]
\centering
\includegraphics[width=3.1in]{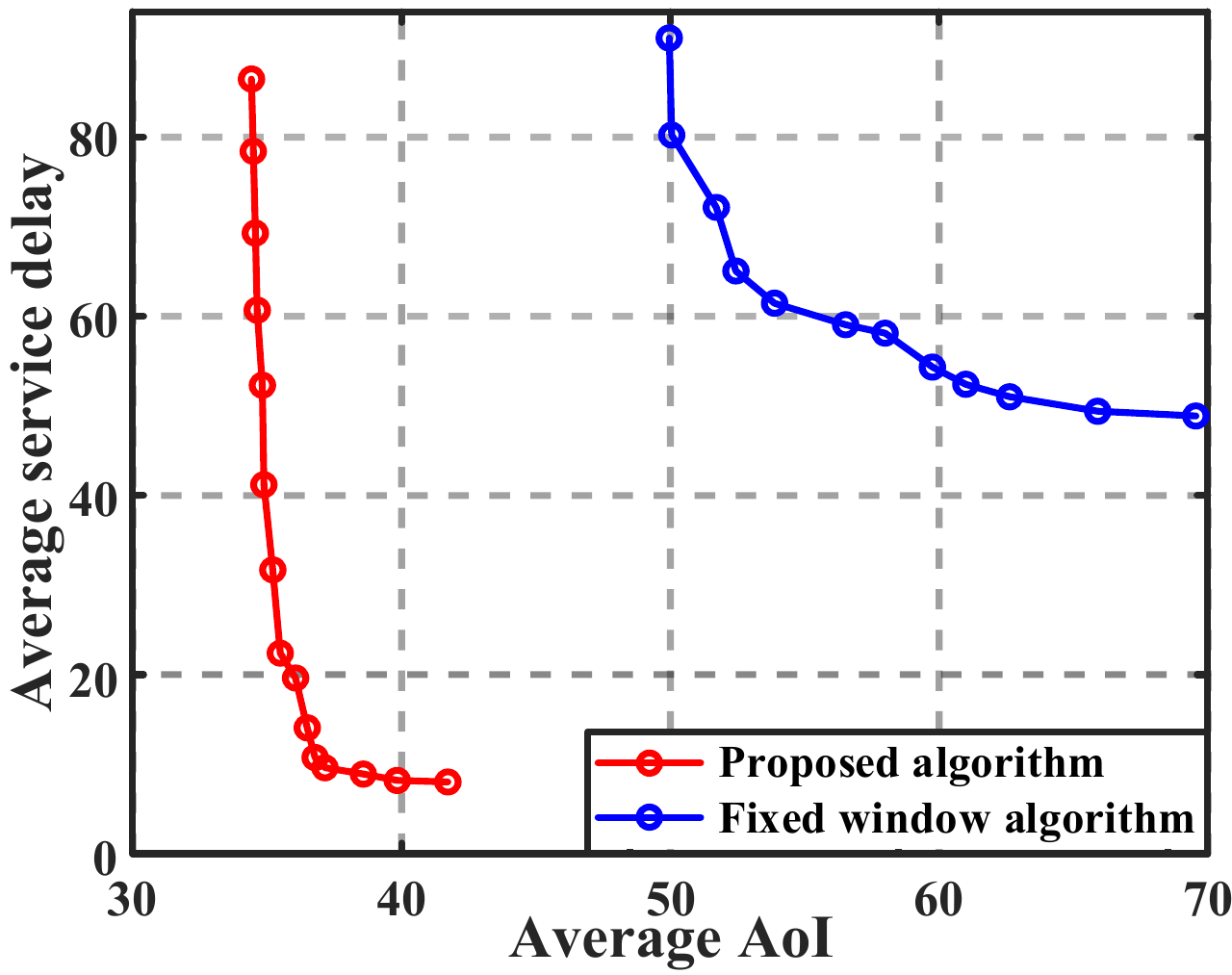}
\caption{Average AoI vs. average service delay for various algorithms in the scenario with $M=10$, $N=40$, $\bar{K}=3$, and the sum arrival rate being 23.}\label{fig3}
\end{figure}

In Fig. \ref{fig1} (a), we evaluate the performances of various algorithms in the scenario with one SN in the MEC network and one slot in each frame, i.e., $M=1$ and $N=1$. We also set $\bar{K}=1$ and $\kappa_1^{\text{UL}}(t)=1$ to ensure a sufficiently large achievable region for the problem. We observe that within the range of $[0,0.42]$ for the arrival rate $\lambda_{1,1}$, the fixed window algorithm consistently outperforms other algorithms by achieving the lowest average value of the objective function \eqref{newnew}, and the DRL algorithm performs comparably to the fixed window algorithm. However, the average value of \eqref{newnew} fails to converge under all of the algorithms when $\lambda_{1,1}$ exceeds $0.42$, which suggests that $\lambda_{1,1}>0.42$ leads to an empty achievable region. Finally, the proposed algorithm has the worst performance, which is reasonable since the condition formulated in section \ref{sec3}, i.e., $\hat{K}+\sum_{k=1}^{\bar{K}}k\lceil\lambda_k\rceil\leq N$, is not satisfied in this scenario. In Fig. \ref{fig1} (b), we investigate the scenario with 10 SNs in the MEC network and one slot in each frame and illustrate the relationship between the sum arrival rate, i.e., $\sum_{m=1}^M\sum_{k=1}^{\bar{K}}k\lambda_{m,k}$, and the average value of \eqref{newnew}. It is observed that both the fixed window and DRL algorithms exhibit promising convergence performances as in the previous case, while the proposed algorithm still does not perform well since the condition $\hat{K}+\sum_{k=1}^{\bar{K}}k\lceil\lambda_k\rceil\leq N$ is not satisfied.

In Fig. \ref{fig2} and Fig. \ref{fig3}, we evaluate the performances of various algorithms in the scenario with 10 SNs in the MEC network and 40 slots in each frame, i.e., $M=10$ and $N=40$. Additionally, we set $\bar{K}=3$. In this particular scenario, both the proposed algorithm and $\pi_s(\bm{\lambda})$ are applicable since the condition $\hat{K}+\sum_{k=1}^{\bar{K}}k\lceil\lambda_k\rceil\leq N$ is satisfied. However, the DRL algorithm is not applicable in this scenario due to the large action space, which has a cardinality of $(41\times21\times14\times2)^{10}$. Fig. \ref{fig2} (a) shows the relationship between the sum arrival rate and the average value of \eqref{newnew}. We observe that the proposed algorithm consistently achieves significantly lower values of \eqref{newnew} compared to other algorithms, especially in scenarios with large sum arrival rates. This demonstrates the ability of the proposed algorithm to efficiently handle the scenarios with heavy requests. Moreover, we observe that all the algorithms achieve large average value of \eqref{newnew} when the sum arrival rate exceeds 23, indicating that the achievable region is empty beyond this threshold. Fig. \ref{fig2} (b) illustrates the achieved average AoI and service delay under different algorithms, where the average AoI and service delay under the proposed algorithm exhibit stable growth as the sum arrival rate increases. This stability demonstrates the robustness of the proposed algorithm with respect to variations in the sum arrival rate. Fig. \ref{fig2} (c) illustrates the slot utility under different algorithms. We observe that the fixed window algorithm occupies all slots within each frame all the time, whereas the proposed algorithm has an increasing slot utility rate as the sum arrival rate grows and achieves full slot utility rate when the value of sum arrival rate is sufficiently large. This indicates that the proposed algorithm can achieve the same average value of \eqref{newnew} as the other algorithms while utilizing fewer slots in each frame. Moreover, we observe that the proposed algorithm achieves full slot utility rate when the sum arrival rate is around 23, which is the threshold where the achievable region is empty. This suggests that we can use the linearity of the slot utility rate with respect to the sum arrival rate in the proposed algorithm to approximate the threshold for the sum arrival rate. Finally, we consider the scenario with the sum arrival rate being 23, vary the value of $V$, and plot the corresponding AoI-delay tradeoff curve in Fig. \ref{fig3}. The results show that the proposed algorithm achieves a substantially lower tradeoff curve, indicating that it can always achieve lower average AoI or lower average service delay than the fixed window algorithm.
\begin{figure*}[htb]
\centering
\includegraphics[width=6.3in]{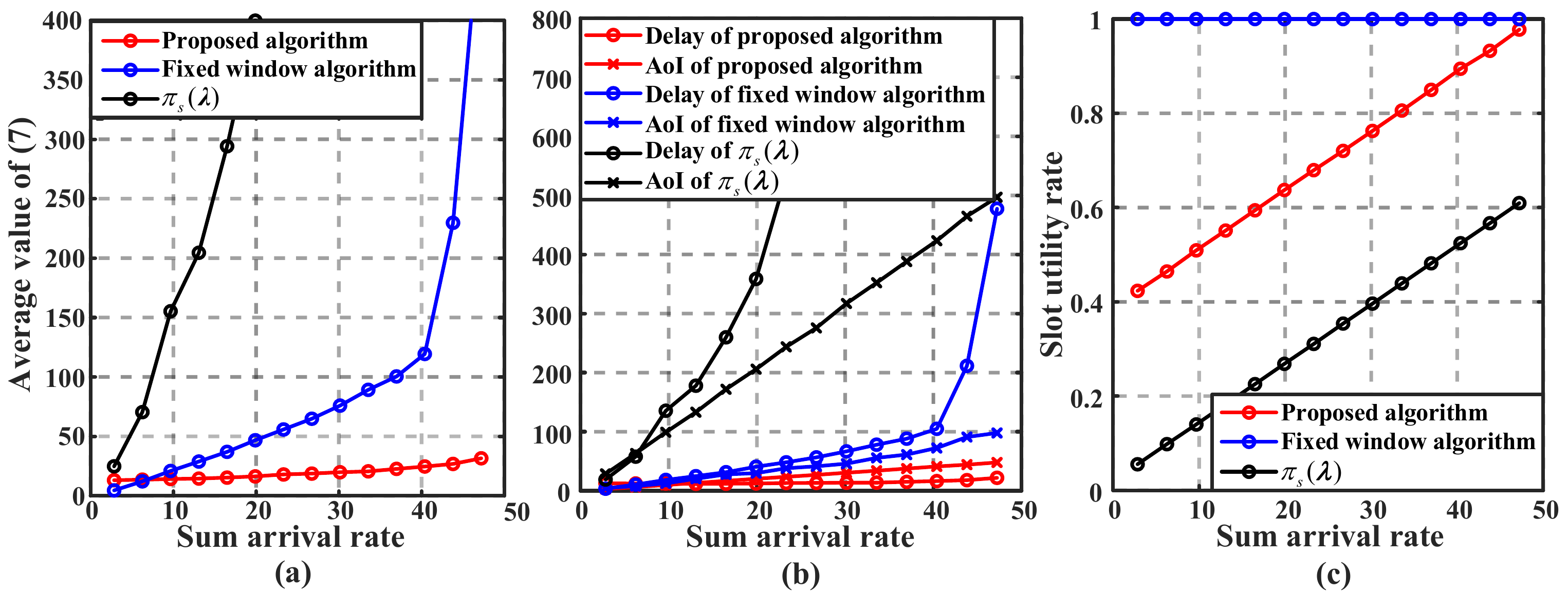}
\caption{Performances of various algorithms in the scenario with $M=20$, $N=80$, and $\bar{K}=3$. (a) Sum arrival rate vs. average value of \eqref{newnew}; (b) Sum arrival rate vs. delay/AoI; (c) Sum arrival rate vs. slot utility rate.}\label{fig4}
\end{figure*}
\begin{figure}[htb]
\centering
\includegraphics[width=3.1in]{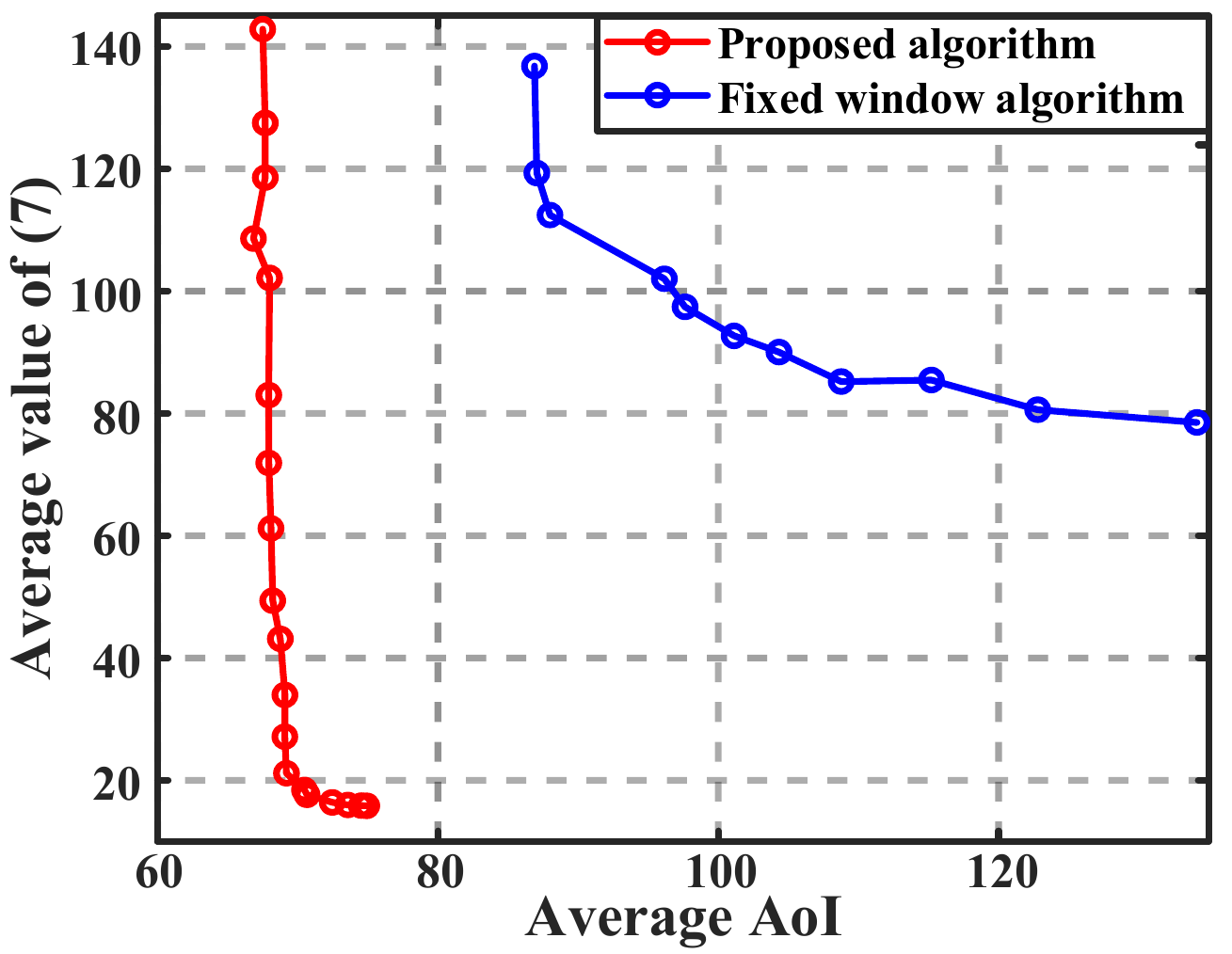}
\caption{Average AoI vs. average service delay for various algorithms in the scenario with $M=20$, $N=80$, $\bar{K}=3$, and the sum arrival rate being 42.}\label{fig5}
\end{figure}

In Fig. \ref{fig4} and Fig. \ref{fig5}, we evaluate the algorithm performances in a more complex scenario with 20 SNs in the MEC network and 80 slots in each frame. We observe that the proposed algorithm can effectively handle the cases with heavy requests, as demonstrated in Fig. \ref{fig4} (a) and Fig. \ref{fig4} (b), by achieving a much lower average value of \eqref{newnew} and stabler average AoI and average service delay than other algorithms. Furthermore, by combining the results of Fig. \ref{fig4} (a) and Fig. \ref{fig4} (c), we again demonstrate that the proposed algorithm can achieve the same average value of \eqref{newnew} as the other algorithms while utilizing fewer slots in each frame. Finally, Fig. \ref{fig5} shows that the proposed algorithm can achieve a better AoI-delay tradeoff than the fixed window algorithm in complex MEC networks.

\section{Conclusions}\label{sec7}
This paper considers the AoI-delay tradeoff in a discrete-time MEC network with multiple SNs in the network and multiple slots in one frame. We formulate the problem as a sequential decision-making problem and derive a superset and a subset of the achievable region using rate stability theorem and a novel stochastic policy. We also derive a sufficient condition for checking the solution's existence by analyzing the features of the subset. To optimize the average AoI and average service delay jointly, we propose a mixed-order drift-plus-penalty algorithm that uses DP to maximize the summation of a linear Lyapunov drift, a quadratic Lyapunov drift, and a penalty term. The proposed algorithm can optimize the objective function with non-linear terms. Theoretical analysis shows that the proposed algorithm achieves an $O(1/V)$ versus $O(V)$ tradeoff for average AoI and average service delay.

\appendices
\setcounter{section}{0}
\section{Proof of Theorem \ref{thm1.1}}\label{app:A}
To prove Theorem \ref{thm1.1}, we first propose a stochastic slot allocation policy $\pi_s(\bm{\lambda})$ and specify its slot allocation method within each frame. Then, based on this policy, we prove that $\bar{\mu}(N)\subseteq\mu(N)$. Finally, we prove that $\lim_{N\rightarrow\infty}{\text{Vol}(\bar{\mu}(N))}/{\text{Vol}(\mu(N))}=1$.

\subsection{Policy $\pi_s(\bm{\lambda})$}
We propose a stochastic policy $\pi_s(\bm{\lambda})$ that allocates slots within each frame using the following two procedures.

\subsubsection{}
In the first procedure, the $1$\ts{st} to the $\hat{K}$\ts{th} slots within each frame are simultaneously allocated to upload a random message out of the $M$ messages and each message is selected for uploading with a probability of $\frac{1}{M}$. In other words, it follows
\begin{align}\label{eq:proc_i}
p_{1:\hat{K},m,\bar{K}+1}(t)=\frac{1}{M},\ \forall m\in\mathcal{M}, t\in\mathbb{Z}_{>0},
\end{align}
where $p_{1:\hat{K},m,\bar{K}+1}(t)$ represents the probability of simultaneously allocating the $1$\ts{st} to the $\hat{K}$\ts{th} slots to upload the $m$\ts{th} message within the $t$\ts{th} frame.

\subsubsection{}
The second procedure consists of $\bar{K}$ steps. In the $k$\ts{th} step, we group the $(\hat{K}+\sum_{k'=1}^{k-1}k'\lceil\lambda_{k'}\rceil+1)$\ts{th} to the $(\hat{K}+\sum_{k'=1}^{k}k'\lceil\lambda_{k'}\rceil)$\ts{th} slots into $\lceil\lambda_{k}\rceil$ sets, each consisting of $k$ slots. Within each set, we utilize all the contained slots to serve one request from either the $(1,k)$\ts{th}, the $(2,k)$\ts{th}, $\cdots$, or the $(M,k)$\ts{th} request queue and the probabilities of serving one request from these request queues are $\frac{\lambda_{1,k}}{\lceil\lambda_k\rceil},\frac{\lambda_{2,k}}{\lceil\lambda_k\rceil},\cdots,\frac{\lambda_{M,k}}{\lceil\lambda_k\rceil}$, respectively, i.e.,
\begin{align}\label{eq:proc_ii}
\begin{split}
&p_{\hat{K}\!+\!\sum_{k'=1}^{k-1}k'\lceil\lambda_{k'}\rceil\!+\!k(n-1)\!+\!1:\hat{K}\!+\!\sum_{k'=1}^{k-1}k'\lceil\lambda_{k'}\rceil\!+\!kn,m,k}(t)\\
=&\frac{\lambda_{m,k}}{\lceil\lambda_k\rceil},
\end{split}
\end{align}
for all $n\in\{1,2,\cdots,\lceil\lambda_k\rceil\}$, $m\in\mathcal{M}$, and $k\in\bar{\mathcal{K}}$, where in the above equality, the notation $p_{\hat{K}+\sum_{k'=1}^{k-1}k'\lceil\lambda_{k'}\rceil+k(n-1)+1:\hat{K}+\sum_{k'=1}^{k-1}k'\lceil\lambda_{k'}\rceil+kn,m,k}(t)$ represents the probability of simultaneously allocating the $(\hat{K}+\sum_{k'=1}^{k-1}k'\lceil\lambda_{k'}\rceil+k(n-1)+1)$\ts{th} to the $(\hat{K}+\sum_{k'=1}^{k-1}k'\lceil\lambda_{k'}\rceil+kn)$\ts{th} slots to serve one request from the $(m,k)$\ts{th} request queue.

In summary, the policy $\pi_s(\bm{\lambda})$ allocates $\hat{K}$ slots for uplink transmissions in the first procedure and $\sum_{k=1}^{\bar{K}}k\lceil\lambda_{k}\rceil$ slots for downlink transmissions in the second procedure. Therefore, to ensure the proper execution of policy $\pi_s(\bm{\lambda})$, the total number of the allocated slots $\hat{K}+\sum_{k=1}^{\bar{K}}k\lceil\lambda_{k}\rceil$ must not exceed the number of slots in one frame, i.e., $N$. In other words, $\bm{\lambda}\in\bar{\mu}(N)$ must hold.

\subsection{Proof of $\bar{\mu}(N)\subseteq\mu(N)$}
Now, we prove that $\bar{\mu}(N)$ is a subset of $\mu(N)$. 

For any $\bm{\lambda}\in\bar{\mu}(N)$, we construct policy $\pi_s(\bm{\lambda})$ based on the aforementioned two procedures. Then, based on \eqref{eq:proc_i}, we have $\mathbb{E}_{\pi_s(\bm{\lambda})}[a_{m,\bar{K}+1}(t)]>0$. By combining this inequality with \eqref{eq:x_m}, we can verify that the average AoI under the policy $\pi_s(\bm{\lambda})$ is finite, i.e, 
$$\lim_{T\rightarrow\infty}\frac{1}{T}\sum_{t=1}^T\sum_{m=1}^M\sum_{k=1}^{\bar{K}}a_{m,k}(t)\left(x_m(t)+1\right)<\infty.$$
Additionally, based on \eqref{eq:proc_ii}, it follows that $a_{m,k}(t)$ is i.i.d. across $t$, and $\mathbb{E}_{\pi_s(\bm{\lambda})}[a_{m,k}(t)]=\lceil\lambda_k\rceil\frac{\lambda_{m,k}}{\lceil\lambda_k\rceil}=\lambda_{m,k}$. By combing these results with the rate stability theorem \cite[Theorem 2.4]{neely_queue}, the process $\{q_{m,k}(t)\}_{t=1}^T$, $m\in\mathcal{M}$ and $k\in\bar{\mathcal{K}}$, is guaranteed to be rate stable under policy $\pi_s(\bm{\lambda})$, i.e.,
\begin{align*}
	\lim_{t\rightarrow 0}\frac{q_{m,k}(t)}{t}=0\ \text{with probability 1},
\end{align*}
which ensures that the average service delay under policy $\pi_s(\bm{\lambda})$ is finite, i.e, 
$$\lim_{T\rightarrow\infty}\frac{1}{T}\sum_{t=1}^T\sum_{m=1}^M\sum_{k=1}^{\bar{K}}\lambda_{m,k}q_{m,k}(t)<\infty.$$ 
Consequently, the objective function \eqref{newnew} of problem {\bf (P1)} under policy $\pi_s(\bm{\lambda})$, which is the summation of the average AoI and the average service delay, is also finite, i.e.,
\begin{align}\label{eq:A_final}
f_{\pi_s(\bm{\lambda})}(\boldsymbol{\lambda},N)<\infty.
\end{align}
Finally, by combining \eqref{eq:A_final} and Definition \ref{defonly}, $\boldsymbol{\lambda}\in\mu(N)$ holds, which implies $\bar{\mu}(N)\subseteq\mu(N)$.

\subsection{Proof of $\lim_{N\rightarrow\infty}{\normalfont\text{Vol}(\bar{\mu}(N))}/{\normalfont\text{Vol}(\mu(N))}=1$} 

First of all, we show that $\hat{\mu}(\hat{N})$ with $\hat{N}\triangleq N-\frac{\bar{K}(\bar{K}+1)}{2}-\hat{K}$ is a subset of $\bar{\mu}(N)$. Specifically, for any $\boldsymbol{\lambda}\in\hat{\mu}(\hat{N})$, we refer the definition in \eqref{def:hat_mu} and obtain
\begin{align}\label{eq:nonono}
	\sum_{m=1}^M\sum_{k=1}^{\bar{K}}k\lambda_{m,k}\leq N-\frac{\bar{K}(\bar{K}+1)}{2}-\hat{K}.
\end{align}
Then, by combing the fact $\lceil\lambda_k\rceil\leq\lambda_k+1$ and inequality \eqref{eq:nonono}, we obtain
\begin{align}\label{eq:appnouse}
\begin{split}
&\hat{K}+\sum_{k=1}^{\bar{K}}k\lceil\lambda_{k}\rceil\leq\hat{K}+\sum_{k=1}^{\bar{K}}k(\lambda_k+1)\\
=&\hat{K}+\frac{\bar{K}(\bar{K}+1)}{2}+\sum_{m=1}^M\sum_{k=1}^{\bar{K}}k\lambda_{m,k}\leq N.
\end{split}
\end{align}
Based on inequality \eqref{eq:appnouse} and the definition of $\bar{\mu}(N)$ in \eqref{def:mu_pi_s}, we have $\boldsymbol{\lambda}\in\bar{\mu}(N)$, which implies that $\hat{\mu}(\hat{N})$ is a subset of $\bar{\mu}(N)$.

Based on the above results and the definition of $\hat{\mu}(N)$ in \eqref{def:hat_mu}, we obtain
\begin{align}\label{eq:suibianba}
	\hat{\mu}(\hat{N})\subseteq\bar{\mu}(N)\subseteq\mu(N)\subseteq\hat{\mu}(N).
\end{align}
Moreover, based on the definition in \eqref{def:hat_mu}, we have
\begin{align*}
&\lim_{N\rightarrow\infty}\frac{\text{Vol}(\hat{\mu}(\hat{N}))}{\text{Vol}(\hat{\mu}(N))}=\lim_{N\rightarrow\infty}\frac{\int_{\boldsymbol{\lambda}\in\hat{\mu}(\hat{N})}\text{d}\boldsymbol{\lambda}}{\int_{\boldsymbol{\lambda}\in\hat{\mu}(N)}\text{d}\boldsymbol{\lambda}}\\
=&\lim_{N\rightarrow\infty}\frac{\frac{1}{(M\bar{K})!}\Pi_{m=1}^{M}\Pi_{k=1}^{\bar{K}}\frac{\hat{N}}{k}}{\frac{1}{(M\bar{K})!}\Pi_{m=1}^{M}\Pi_{k=1}^{\bar{K}}\frac{N}{k}}=1.
\end{align*}
By combining this result with \eqref{eq:suibianba}, we have
\begin{align*}
\lim_{N\rightarrow\infty}\frac{\text{Vol}(\bar{\mu}(N))}{\text{Vol}(\mu(N))}=1,
\end{align*}
which completes the proof.

\section{Proof of Proposition \ref{lemma1}}\label{app:B}
Based on the equalities in \eqref{eq:x_m} and \eqref{eq:q_m_k} and the definitions in \eqref{def:l_x} and \eqref{def:l_q}, it follows
\begin{align}
&L(\bm{x}(t+1))-L(\bm{x}(t))\nonumber\\
=&\sum_{m=1}^M(x_m(t+1)-x_m(t))\nonumber\\
=&\sum_{m=1}^M(1-a_{m,\bar{K}+1}(t)x_m(t)),\label{eq:prf2_0}
\end{align}
\begin{align}
&L(\bm{Q}(t+1))\!-\!L(\bm{Q}(t))\nonumber\\
=&\frac{1}{2}\!\sum_{m=1}^M\!\sum_{k=1}^{\bar{K}}\!\lambda_{m,k}\!\Big(\!\left(\max\{q_{m,k}(t)\!-\!a_{m,k}(t),\!0\}\!+\!c_{m,k}(t)\right)^2\nonumber\\
&-\!q_{m,k}^2(t)\Big)\nonumber\\
\begin{split}\label{eq:prf2_1}
\leq&\frac{1}{2}\sum_{m=1}^M\sum_{k=1}^{\bar{K}}\lambda_{m,k}\Big(c_{m,k}^2(t)\!+\!a_{m,k}^2(t)\\
&-\!2q_{m,k}(t)\left(a_{m,k}(t)\!-\!c_{m,k}(t)\right)\Big).
\end{split}
\end{align}
Inequality \eqref{eq:prf2_1} is obtained by considering the fact that for any $x\geq0$, $y\geq0$, and $z\geq0$, it follows $(\max\{x-y,0\}+z)^2\leq x^2+y^2+z^2+2x(z-y)$.

Next, by plugging the definitions in \eqref{def:drift_q} and \eqref{def:drift_x}, equality \eqref{eq:prf2_0}, and inequality \eqref{eq:prf2_1} into \eqref{eq:drift_pi1}, we obtain
\begin{align}
&\eqref{eq:drift_pi1}\nonumber\\
\begin{split}\label{eq:prf2_2}
\leq&\frac{1}{2}\!\sum_{m=1}^M\!\sum_{k=1}^{\bar{K}}\!\lambda_{m,k}\!\Big(\mathbb{E}_{c_{m,k}(t)}[c_{m,k}^2(t)]\!+\!\mathbb{E}_{\pi}\left[a_{m,k}^2(t)|\bm{x}(t),\!\bm{Q}(t)\right]\\
&-2q_{m,k}(t)(\mathbb{E}_{\pi}[a_{m,k}(t)|\bm{x}(t),\bm{Q}(t)]\!-\!\lambda_{m,k})\Big)\\
&+\!VV_0\sum_{m=1}^M(1\!-\!\mathbb{E}_{\pi}\!\left[a_{m,\bar{K}+1}(t)|\bm{x}(t),\!\bm{Q}(t)\right]\!x_m(t))\\
&+V\sum_{m=1}^M\sum_{k=1}^{\bar{K}}\mathbb{E}_{\pi}\left[a_{m,k}(t)|\bm{x}(t),\bm{Q}(t)\right](x_m(t)\!+\!1).
\end{split}
\end{align}

Then, based on inequality \eqref{con:a_m_k_1}, we have
\begin{align}
\begin{split}\label{eq:prf2_3}
&\frac{1}{2}\sum_{m=1}^M\sum_{k=1}^{\bar{K}}\lambda_{m,k}\mathbb{E}_{\pi}\left[a_{m,k}^2(t)|\bm{x}(t),\bm{Q}(t)\right]\\
\leq&\frac{1}{2}\max_{m\in\mathcal{M},k\in\bar{\mathcal{K}}}\lambda_{m,k}\Big\lceil\frac{N}{k}\Big\rceil^2.
\end{split}
\end{align}

Finally, combing equality \eqref{eq:prf2_2} and inequality \eqref{eq:prf2_3}, it yields
\begin{align*}
&\eqref{eq:drift_pi1}\\
\begin{split}
\leq&\frac{1}{2}\sum_{m=1}^M\sum_{k=1}^{\bar{K}}\lambda_{m,k}\mathbb{E}_{c_{m,k}(t)}[c_{m,k}^2(t)]+\frac{1}{2}\max_{m\in\mathcal{M},k\in\bar{\mathcal{K}}}\lambda_{m,k}\Big\lceil\frac{N}{k}\Big\rceil^2\\
&-\sum_{m=1}^M\sum_{k=1}^{\bar{K}}\lambda_{m,k}q_{m,k}(t)\left(\mathbb{E}_{\pi}\left[a_{m,k}(t)|\bm{x}(t),\bm{Q}(t)\right]-\lambda_{m,k}\right)\\
&-VV_0\sum_{m=1}^M\mathbb{E}_{\pi}[a_{m,\bar{K}+1}(t)|\bm{x}(t),\bm{Q}(t)]x_m(t)+VV_0M\\
&+V\sum_{m=1}^M\sum_{k=1}^{\bar{K}}\mathbb{E}_{\pi}\left[a_{m,k}(t)|\bm{x}(t),\bm{Q}(t)\right](x_m(t)+1)
\end{split}\\
=&\eqref{eq:drift_pi2},
\end{align*}
which completes the proof.

\section{Proof of Proposition \ref{prop5}}\label{app:B2}
First, since policy $\pi_m$ is the solution to problem {\bf (P3)}, we have
\begin{align}\label{eq:drift_pi_d_king}
\eqref{eq:drift_pi2}|_{\pi=\pi_m}\leq\eqref{eq:drift_pi2}|_{\pi=\pi_0}, \ \forall\pi_0\in\Pi,
\end{align}
where $\Pi$ is defined as the set containing all feasible slot allocation policies. 

Next, since $\bm{\lambda}+\epsilon(\bm{\lambda})\cdot1^{M\times\bar{K}}\in\bar{\mu}(N)$ holds, based on the definition of $\bar{\mu}(N)$ in \eqref{def:mu_pi_s}, we have
\begin{align}\label{eq:lambda}
\pi_s(\bm{\lambda}+\epsilon\cdot1^{M\times\bar{K}})\in\Pi,\ \forall\epsilon\in[0,\epsilon(\bm{\lambda})].
\end{align}
By combing Proposition \ref{lemma1}, equality \eqref{eq:drift_pi_d_king}, and \eqref{eq:lambda}, we have
\begin{align}\label{eq:drift_pi_d1}
\eqref{eq:drift_pi1}|_{\pi=\pi_m}\leq\eqref{eq:drift_pi2}|_{\pi=\pi_m}\leq\eqref{eq:drift_pi2}|_{\pi=\pi_s(\bm{\lambda}+\epsilon\cdot1^{M\times\bar{K}})}.
\end{align}

Then, based on Appendix \ref{app:A}, it follows
\begin{align}\label{eq:pi_s}
\mathbb{E}_{\pi_s(\bm{\lambda}+\epsilon\cdot1^{M\times\bar{K}})}\!\left[a_{m,k}(t)|\bm{x}(t),\bm{Q}(t)\right]\!=\!\lambda_{m,k}\!+\!\epsilon,
\end{align}
for all $m\in\mathcal{M}$, $k\in\bar{\mathcal{K}},t\in\mathbb{Z}_{>0}$, and $\epsilon\in[0,\epsilon(\bm{\lambda})]$. By plugging \eqref{eq:pi_s} into \eqref{eq:drift_pi_d1}, we have
\begin{align}\label{eq:drift_pi_d2}
\begin{split}
\eqref{eq:drift_pi1}|_{\pi=\pi_m}\leq&C+VV_0M+V\sum_{m=1}^M\sum_{k=1}^{\bar{K}}(\lambda_{m,k}+\epsilon)\\
&-\!\epsilon\!\sum_{m=1}^M\!\sum_{k=1}^{\bar{K}}\lambda_{m,k}q_{m,k}(t)\!-\!\frac{VV_0}{M}\sum_{m=1}^Mx_m(t)\\
&+V\sum_{m=1}^M\sum_{k=1}^{\bar{K}}(\lambda_{m,k}+\epsilon)x_m(t).
\end{split}
\end{align}

Finally, denote the distributions of $\bm{x}(t)$ and $\bm{Q}(t)$ under policy $\pi_m$ as $\pi_m(\bm{x}(t))$ and $\pi_m(\bm{Q}(t))$, respectively. Taking expectation for \eqref{eq:drift_pi_d2} over policy $\pi_m$, it yields
\begin{align*}
\mathbb{E}_{\pi_m}\![\eqref{eq:drift_pi1}|_{\pi=\pi_m}]=\mathbb{E}_{\bm{x}(t)\sim \pi_m(\bm{x}(t))\atop\bm{Q}(t)\sim\pi_m(\bm{Q}(t))}\left[\eqref{eq:drift_pi1}|_{\pi=\pi_m}\right]\leq \eqref{eq:drift_pi_d3},
\end{align*}
which completes the proof.

\section{Proof of Theorem \ref{thm5.1}}\label{app:C}
First, by summing up $\mathbb{E}_{\pi_m}[\eqref{eq:drift_pi1}|_{\pi=\pi_m}]$ over $t\in\{1,2\cdots,$ $T\}$, we have
\begin{align}
&\sum_{t=1}^T\mathbb{E}_{\pi_m}[\eqref{eq:drift_pi1}|_{\pi=\pi_m}]\nonumber\\
\begin{split}
=&\!\!\sum_{t=1}^T\!\mathbb{E}\!_{\bm{x}(t)\!\sim\!\pi_m(\!\bm{x}(t)\!)\atop\bm{Q}(t)\!\sim\!\pi_m(\!\bm{Q}\!(t)\!)}\!\!\mathbb{E}_{\pi_m,c_{m,k}(t)}\!\!\left[L(\bm{Q}(t\!+\!1))\!-\!L(\bm{Q}(t))|\bm{x}(t)\!,\!\bm{Q}(t)\!\right]\nonumber\\
&+\!VV_0\sum_{t=1}^T\mathbb{E}_{\bm{x}(t)\sim\pi_m(\bm{x}(t))\atop\bm{Q}(t)\sim\pi_m(\bm{Q}(t))}\mathbb{E}_{\pi_m,c_{m,k}(t)}\Big[L(\bm{x}(t+1))\\
&-L(\bm{x}(t))\Big|\bm{x}(t),\bm{Q}(t)\Big]\nonumber\\
&+V\sum_{t=1}^T\sum_{m=1}^M\sum_{k=1}^{\bar{K}}\mathbb{E}\!_{\bm{x}(t)\sim\pi_m(\bm{x}(t))\atop\bm{Q}(t)\sim\pi_m(\bm{Q}(t))}\!\Big[\mathbb{E}_{\pi_m}\!\left[a_{m,k}(t)|\bm{x}(t),\bm{Q}(t)\right]\\
&(x_m(t)\!+\!1)\Big]\nonumber
\end{split}\\
\begin{split}\label{eq:drift_pi_d4}
=&\sum_{t=1}^T\mathbb{E}_{\pi_m,c_{m,k}(t)}\left[L(\bm{Q}(t+1))-L(\bm{Q}(t))\right]\\
&+VV_0\sum_{t=1}^T\mathbb{E}_{\pi_m,c_{m,k}(t)}\left[L(\bm{x}(t+1))-L(\bm{x}(t))\right]\\
&+V\sum_{t=1}^T\sum_{m=1}^M\sum_{k=1}^{\bar{K}}\mathbb{E}_{\pi_m,c_{m,k}(t)}\!\left[a_{m,k}(t)(x_m(t)+1)\right]
\end{split}\\
\begin{split}\label{eq:drift_pi_d5}
=&\mathbb{E}_{\pi_m,c_{m,k}(t)}\left[L(\bm{Q}(T+1))-L(\bm{Q}(1))\right]\\
&+VV_0\mathbb{E}_{\pi_m,c_{m,k}(t)}\left[L(\bm{x}(T+1))-L(\bm{x}(1))\right]\\
&+V\mathbb{E}_{\pi_m,c_{m,k}(t)}\!\left[\sum_{t=1}^T\sum_{m=1}^M\sum_{k=1}^{\bar{K}}\!a_{m,k}(t)(x_m(t)\!+\!1)\right]\!,
\end{split}
\end{align}
where equality \eqref{eq:drift_pi_d4} is obtained by using the law of iterated expectation \cite{neely_queue}.

Next, by summing up \eqref{eq:drift_pi_d3} over $t\in\{1,2,\cdots,T\}$, we have
\begin{align}
&\sum_{t=1}^T\eqref{eq:drift_pi_d3}\nonumber\\
\begin{split}\label{eq:drift_pi_d6}
=&T\left(C+V\Big(V_0M+\sum_{m=1}^M\sum_{k=1}^{\bar{K}}(\lambda_{m,k}+\epsilon)\Big)\right)\\
&-\epsilon\mathbb{E}_{\pi_m,c_{m,k}(t)}\Bigg[\sum_{t=1}^T\sum_{m=1}^M\sum_{k=1}^{\bar{K}}\lambda_{m,k}q_{m,k}(t)\Bigg]\\
&-\!V\!\mathbb{E}_{\pi_m,c_{m,k}(t)}\!\!\left[\sum_{t=1}^T\!\sum_{m=1}^M\!\sum_{k=1}^{\bar{K}}\!\left(\!\frac{V_0}{M\bar{K}}\!-\!(\lambda_{m,k}+\epsilon)\!\!\right)x_m(t)\right].
\end{split}
\end{align}

Then, based on the fact $\mathbb{E}_{\pi_m}\![\eqref{eq:drift_pi1}|_{\pi=\pi_m}]\leq$\eqref{eq:drift_pi_d3}, we have
\begin{align}\label{eq:drift_pi_dend}
\lim_{T\rightarrow\infty}\frac{\eqref{eq:drift_pi_d5}}{T}\leq\lim_{T\rightarrow\infty}\frac{\eqref{eq:drift_pi_d6}}{T}.
\end{align}

Finally, we fix $V_0$ as
\begin{align}\label{eq:V_0}
V_0=M\bar{K}\max_{m\in\mathcal{M},k\in\bar{\mathcal{K}}}\left(\lambda_{m,k}+\epsilon(\bm{\lambda}\right))
\end{align}
and it follows
\begin{align}\label{eq:V_0_0}
&\frac{V_0}{M\bar{K}}-(\lambda_{m,k}+\epsilon)\geq 0.
\end{align}
By plugging \eqref{eq:V_0_0} into \eqref{eq:drift_pi_dend}, we have
\begin{align*}
&\lim_{T\rightarrow\infty}\frac{1}{T}\mathbb{E}_{\pi_m,c_{m,k}(t)}\left[L(\bm{Q}(T+1))-L(\bm{Q}(1))\right]\\
&+\lim_{T\rightarrow\infty}\frac{1}{T}VV_0\mathbb{E}_{\pi_m,c_{m,k}(t)}\left[L(\bm{x}(T+1))-L(\bm{x}(1))\right]\\
&+\lim_{T\rightarrow\infty}\frac{1}{T}V\mathbb{E}_{\pi_m,c_{m,k}(t)}\!\left[\sum_{t=1}^T\sum_{m=1}^M\sum_{k=1}^{\bar{K}}a_{m,k}(t)(x_m(t)\!+\!1)\right]\!\\
\leq&C\!+\!V\Big(\!\max_{m\in\mathcal{M},k\in\bar{\mathcal{K}}}\!\left(\lambda_{m,k}\!\!+\!\epsilon(\bm{\lambda}\right))M^2\bar{K}\!+\!\!\sum_{m=1}^M\!\sum_{k=1}^{\bar{K}}(\lambda_{m,k}+\epsilon)\Big)\\
&-\epsilon\lim_{T\rightarrow\infty}\frac{1}{T}\mathbb{E}_{\pi_m,c_{m,k}(t)}\Bigg[\sum_{t=1}^T\sum_{m=1}^M\sum_{k=1}^{\bar{K}}\lambda_{m,k}q_{m,k}(t)\Bigg],
\end{align*}
which induces that the average AoI of the requests satisfies
\begin{align}
\begin{split}
&\frac{1}{\sum_{m=1}^M\!\sum_{k=1}^{\bar{K}}\!\lambda_{m,k}}\!\lim_{T\rightarrow\infty}\frac{1}{T}\mathbb{E}_{\pi_m,c_{m,k}(t)}\!\Bigg[\sum_{t=1}^T\!\sum_{m=1}^M\!\sum_{k=1}^{\bar{K}}\nonumber\\
&a_{m,k}(t)(x_m(t)\!+\!1)\Bigg]\nonumber
\end{split}\\
\begin{split}\label{eq:bound_aoi}
\leq&\frac{1}{\sum_{m=1}^M\sum_{k=1}^{\bar{K}}\lambda_{m,k}}\Bigg(\max_{m\in\mathcal{M},k\in\bar{\mathcal{K}}}\left(\lambda_{m,k}+\epsilon(\bm{\lambda}\right))M^2\bar{K}\\
&+\sum_{m=1}^M\sum_{k=1}^{\bar{K}}(\lambda_{m,k}+\epsilon)+\frac{C}{V}\Bigg),\forall\epsilon\in[0,\epsilon(\bm{\lambda})],
\end{split}
\end{align}
and the average service delay of the requests satisfies
\begin{align}
\begin{split}
&\frac{1}{\sum_{m=1}^M\!\sum_{k=1}^{\bar{K}}\!\lambda_{m,k}}\lim_{T\rightarrow\infty}\frac{1}{T}\mathbb{E}_{\pi_m,c_{m,k}(t)}\Bigg[\sum_{t=1}^T\sum_{m=1}^M\sum_{k=1}^{\bar{K}}\lambda_{m,k}\\
&q_{m,k}(t)\Bigg]+1\nonumber
\end{split}\\
\begin{split}\label{eq:bound_delay}
\leq&\frac{1}{\epsilon\sum_{m=1}^M\sum_{k=1}^{\bar{K}}\lambda_{m,k}}\Bigg(\Big(\max_{m\in\mathcal{M},k\in\bar{\mathcal{K}}}\left(\lambda_{m,k}\!+\!\epsilon(\bm{\lambda}\right))M^2\bar{K}\\
&+\sum_{m=1}^M\sum_{k=1}^{\bar{K}}(\lambda_{m,k}+\epsilon)\Big)V+C\Bigg)+1,\ \forall\epsilon\in(0,\epsilon(\bm{\lambda})].
\end{split}
\end{align}

Since inequality \eqref{eq:bound_aoi} holds for all $\epsilon\in[0,\epsilon(\bm{\lambda})]$, we set $\epsilon$ as $0$ and derive inequality \eqref{eq:aoi_final}. Similarly, in inequality \eqref{eq:bound_delay}, we set $\epsilon$ as $\epsilon(\bm{\lambda})$ and derive inequality \eqref{eq:delay_final}.


\bibliographystyle{IEEEtran}
\bibliography{refs}

\end{document}